\documentclass[conference]{IEEEtran}
\IEEEoverridecommandlockouts

\usepackage{amsmath,amssymb,amsfonts}

\usepackage{graphicx}

\usepackage[noend]{algpseudocode}
\usepackage{amsmath}
\usepackage{enumitem}
\usepackage{textcomp}
\usepackage{xcolor}
\usepackage[utf8]{inputenc}
\usepackage{amsmath}
\usepackage{booktabs,tabularx}
\usepackage[noadjust]{cite}
\usepackage{authblk}
\usepackage{subcaption}
\usepackage[utf8]{inputenc}
\usepackage[english]{babel}
\usepackage{amsthm}
\usepackage{mathtools}
\usepackage{multirow}
\usepackage[square,sort,comma,numbers]{natbib}
\usepackage{multicol}

\usepackage{url}
\usepackage{caption}
\usepackage{subcaption}
\usepackage{tabularx}
\usepackage{verbatim}
\usepackage[linesnumbered,ruled,vlined]{algorithm2e} 
\usepackage[export]{adjustbox}

\usepackage{amsmath}

\def\BibTeX{{\rm B\kern-.05em{\sc i\kern-.025em b}\kern-.08em
    T\kern-.1667em\lower.7ex\hbox{E}\kern-.125emX}}
    
\begin{document}

\title{Graph Convolutional Reinforcement Learning for Collaborative Queuing Agents}
\author[1]{Hassan Fawaz}
\author[2]{Julien Lesca}
\author[2]{Pham Tran Anh Quang}
\author[2]{Jérémie Leguay}
\author[1]{Djamal Zeghlache}
\author[2]{Paolo Medagliani}
\affil[1]{T\'{e}l\'{e}com SudParis, Institut Polytechnique de Paris, Evry, France}
\affil[2]{Huawei Technologies Ltd., Paris Research Center, France}

\newcommand{\jleg}[1]{\textcolor{red}{#1}}
\newcommand{\hf}[1]{\textcolor{blue}{#1}}
\newcommand{\dz}[1]{\textcolor{violet}{#1}}
\newcommand{\pmed}[1]{\textcolor{orange}{#1}}
\newcommand{\jles}[1]{\footnote{\textcolor{green}{#1}}}
\newcommand{\qp}[1]{\textcolor{brown}{#1}}
\maketitle

\thispagestyle{plain}
\pagestyle{plain}
\setlength{\textfloatsep}{2pt}
\begin{abstract}
In this paper, we explore the use of  multi-agent deep learning as well as learning to cooperate principles to meet stringent service level agreements, in terms of throughput and end-to-end delay, for a set of classified network flows. We consider
agents built on top of a weighted fair queuing algorithm that continuously set weights for three flow groups: gold, silver, and bronze. We rely on a novel graph-convolution based, multi-agent reinforcement learning approach known as DGN. As benchmarks, we propose centralized and distributed deep Q-network approaches and evaluate their performances in different network, traffic, and routing scenarios, highlighting the effectiveness of our proposals and the importance of agent cooperation. We show that our DGN-based approach meets stringent throughput and delay requirements across all scenarios.
\end{abstract}

\section{Introduction}
\indent  
Traffic scheduling is key to control how  bandwidth is shared among different applications and in particular, to  satisfy Service Level Agreements (SLA) of applications in terms of throughput, delay, loss and jitter. In typical Software-Defined Wide Area Networks (SD-WAN) architectures~\cite{yang2019software}, a centralized controller maintains a set of policies deployed at edge routers that interconnect multiple sites (enterprise branches, data centers). Each edge router is configured to send traffic to its peers over several transport networks ($e.g.$, private lines based on MPLS or  cheaper broadband internet connections). Typically, these routers are responsible for applying routing and queuing policies to meet SLA requirements in terms of end-to-end Quality of Service (QoS), security, etc. At a slow pace, the controller optimizes policies, while edge devices make real-time decisions.

Several solutions~\cite{yang2019software} have been proposed for the dynamic selection of paths in WAN networks to satisfy SLA requirements. The general idea is to compare the quality of paths with application requirements and update the path selection strategy inside routers when needed. Beyond path selection, a number of adaptive queuing and Active Queue Management (AQM) techniques~\cite{aqmsurvey} have been proposed to help sustain delay and throughput requirements. In particular, the dynamic adaptation of scheduling parameters, such as the weights in Adaptive Weighted Fair Queuing (AWFQ)~\cite{FRANTTI200911390, sayenko2006comparison, homg2001adaptive}, has been shown to significantly improve  performance.
Nonetheless,  existing mechanisms are local and work at the level of individual routers in the network, without trying to explicitly cooperate to globally improve the QoS. In~\cite{HUSSAIN20031143}, for instance, an agent at the destination informs the source node of delay limit violations, so the upstream agent adjusts its queuing weight, but there is no cooperation or sharing of information across agents.
In our work, we design a multi-agent system based on deep reinforcement learning with the objective of improving queue management in networks.

\indent To this end, we propose a set of Deep Reinforcement Learning (DRL) algorithms that optimize queuing parameters to meet SLA requirements. We consider 
a typical SD-WAN scenario in which routers deal with an array of classified flow groups with different requirements in terms of throughput and latency.
A WFQ approach is set up to control how each flow group is served at ingress nodes. 
Our DRL algorithms are embedded into agents controlling WFQ 
weights for each flow group
depending on the traffic and network status at hand. The delay on each path, as well as the eventual achieved throughput by the flows, depends on the interfering traffic present in other queues and on other paths. This necessitates dynamically tuning the weights and motivates cooperation between agents managing the different nodes, $i.e.,$ routers. While closed-form expressions for WFQ~\cite{homg2001adaptive} can be used to tune weights locally, a machine learning approach can better adapt to realistic traffic patterns and generalize to the case where multiple agents are interfering ($e.g.$, sending traffic over the same links) and end-to-end QoS requirements must be met.\\
\indent We utilize a multi-agent approach to tackle the problem. In Multi-Agent Reinforcement Learning (MARL), multiple methods exist to govern agent communication and cooperation. The MARL system could be completely centralized, fully distributed, or semi-distributed. In a centralized MARL system, all the agents act as one, sharing the same environment, states, actions and rewards. In a distributed one the agents are completely independent, and in a semi-distributed MARL architecture, these distributed agents are able to communicate and cooperate.\\
\indent In this paper, our main contribution is the application of a graph convolutional reinforcement learning (DGN) approach to the multi-agent smart queuing problem. DGN is a semi-distributed approach to MARL in which the collaboration between agents can be parameterized and learned. In addition, as benchmark solutions covering other MARL architectures, we propose two Multi-Agent Deep Q-Learning (MADQN) solutions based on DQN~\cite{mnih2015human}. One is completely centralized and the other fully distributed.\\
\indent In our work, we discuss how both the DGN and MADQN agents learn. We detail their observations, actions, rewards, and the extent of their cooperation across the different considered approaches. We perform packet-level simulations in ns-3~\cite{ns3} and compare our proposals against traditional priority queuing (PQ), in both SD-WAN and classic network topologies (Abilene). We show that our proposals are better suited to deal with classified traffic than PQ. While DGN is always capable of meeting the required throughput and delay demands, we illustrate how the lack of agent cooperation in the distributed MADQN approach can cause the latter to falter in convoluted scenarios. And while the centralized approach to MADQN can meet the set objectives, we show that DGN can do it without the need for a centralized setting and with negligible overhead during execution. \\
\indent Section~\ref{related} of this paper describes the related works in the state-of-the-art. Section~\ref{archi} discusses the system architecture, including our SD-WAN use case and the WFQ approach we build our agents upon. Section~\ref{dgn} introduces our graph convolutional reinforcement learning based proposal for smart queue management. Section~\ref{dqn} details both our centralized and distributed deep Q-learning approaches to the smart queuing problem. Section~\ref{results} presents the simulation results and analysis, while Section~\ref{conclusion} concludes this paper.
\section{Related Works}\label{related}
\indent In this section, we discuss the related works on smart queuing and the utilization of deep reinforcement learning (DRL) in network management. In this context, we first focus on active queue management and multi-path traffic engineering.

\begin{figure}[t]
	\begin{center}
		\includegraphics[width=\linewidth]{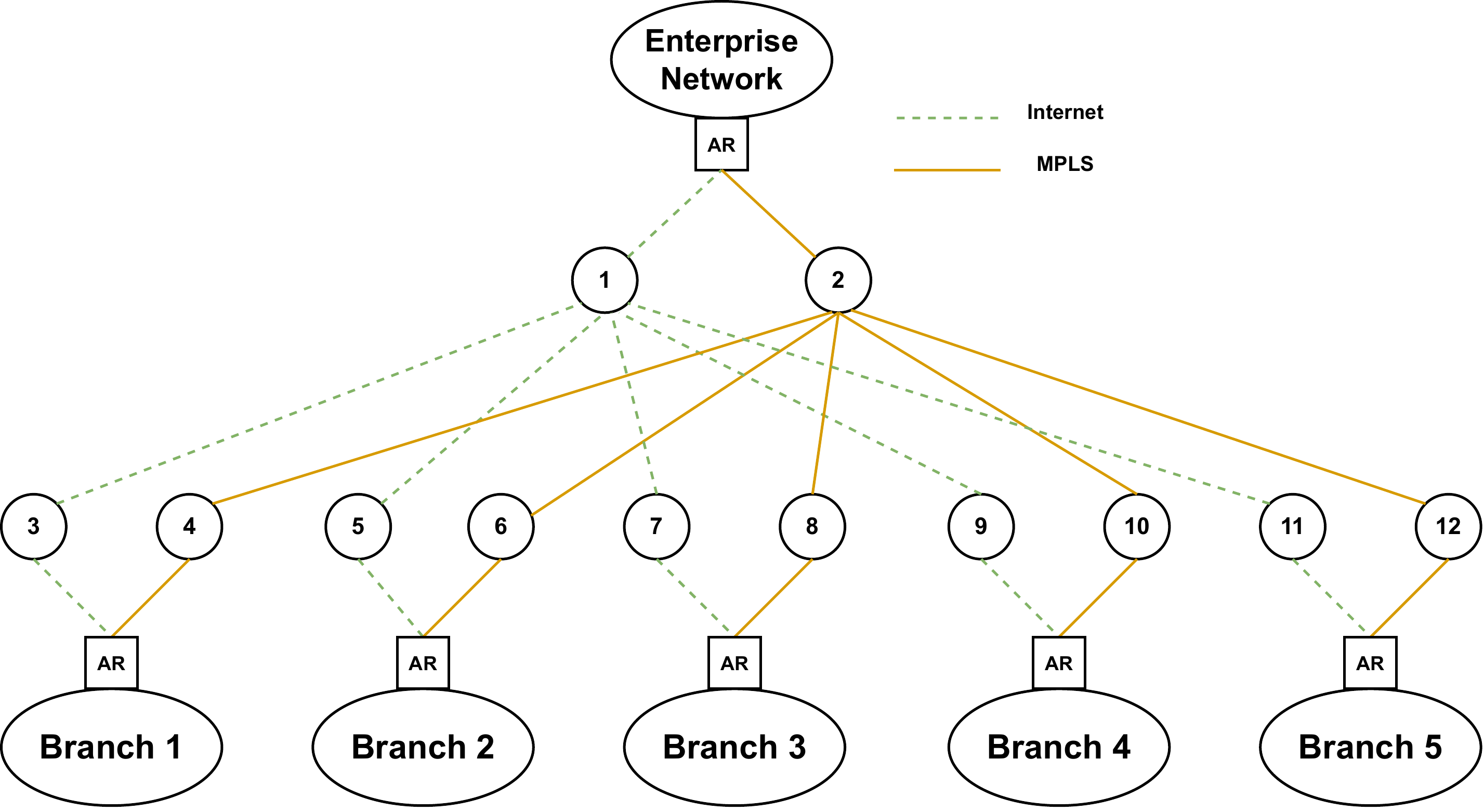}
		\caption{SD-WAN network with 5 branches}
		\label{fig:nettopology}
	\end{center}
\end{figure}
\indent In terms of what we aim to accomplish in this work, the paper of Kim and Eng~\cite{kim2019deep} is the closest. The authors propose a DQN based AQM algorithm in a single-agent environment, wherein the agent decides which packets to serve from the queue and which ones to drop. Other queue-based DRL usages can be seen in the paper of Balasubramanian et al.~\cite{balasubramanian2020reinforcing}, where the agents decide which request traffic instances are to be served first, and in the work of Bachl, Fabini and Zseby~\cite{LFQ}, where they are tasked with finding the optimal buffer sizes.\\
\indent DRL approaches in the domain of multi-path traffic engineering in general and multi-path TCP specifically are also popular. Rosello~\cite{rosello2019multi} proposed a DQN agent with the purpose of selecting the optimal paths for MPTCP, while Liao et al.~\cite{liao2020precise} used an actor-critic framework to the same end. Finally, Houdi et al.~\cite{Houidi2022} proposed a multi-agent actor-critic framework to perform path selection and optimize quality of experience.\\
\begin{figure}[t]
	\begin{center}
		\includegraphics[width=\linewidth]{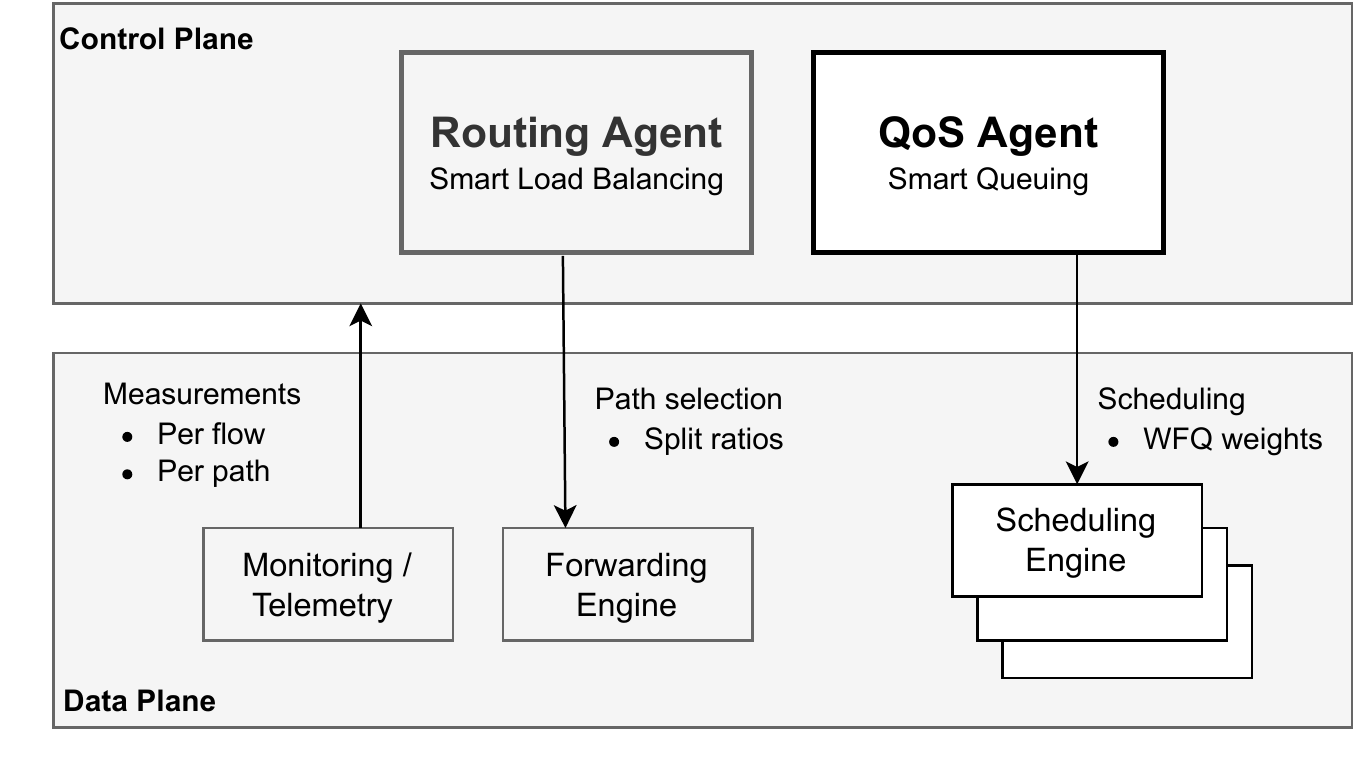}
		\caption{Access router architecture}
		\label{fig:architecture}
	\end{center}
\end{figure}
\indent In this paper, we propose a graph convolutional reinforcement learning multi-agent approach for optimal weight selection in a network using WFQ schedulers. The objective is to meet delay and throughput requirements for a set of classified network flows. Originally proposed by Jiang et al.~\cite{DGN2018}, DGN aims at learning how agents cooperate in a MARL environment. It uses attention~\cite{vaswani2017attention} and adjacency matrices to extract relevant features and relay important information where needed. With respect to the state-of-the-art on cooperation in multi-agent deep reinforcement learning, DGN utilizes attention mechanisms similar to those proposed by Jiang \& Lu~\cite{ATOC}, whilst avoiding its full-scale communication. It uses parameter sharing as done in the proposal by Zhang et al.~\cite{zhang2018fully}, but without assuming a fully observable environment. And finally, while DGN was not the first proposal to utilize a graph convolutional network, Malysheva et al.~\cite{malysheva2018deep}, Agarwal et al.~\cite{agarwal2019learning}, among others, it does so in a partially observable environment whilst allowing for a dynamic adjacency of agents.
Our paper goes beyond state-of-the-art by considering a multi-agent architecture based on reinforcement learning algorithms, such as DGN, for the adaptive tuning of queuing parameters to meet SLA requirements.
\section{System Architecture}\label{archi}
\indent We consider a semi-distributed architecture where edge devices are controlling traffic based on real-time measurements using local agents sharing some information with their peers. The agents are centrally trained, but their execution is done in a distributed manner. In this section, 
we detail the architecture of the SD-WAN use case that we focus on, and afterwards we discuss the scheduling approach on which we build our reinforcement learning proposals.
\subsection{SD-WAN Use Case}
\indent
Figure~\ref{fig:nettopology} presents a typical SD-WAN use case where an enterprise network headquarters (HQ) and five remote branches are interconnected by MPLS and broadband internet connections controlled by third-party operators. A controller is placed at the headquarter site and access routers (ARs) are responsible for the interconnection. Flows issued by user applications are grouped into \emph{flow groups} that correspond to traffic classes with different SLA requirements. A typical traffic scenario includes gold, silver, and bronze groups for multimedia, business critical, and non-critical applications, respectively. \\
\indent The system architecture is split into two control entities operating at two
different time scales.
In a slow control loop, the global controller (at the headquarter site) updates policies and communicates them to edge devices ($i.e.,$ AR devices). In a fast control loop, devices take tactical decisions to follow the evolution of traffic and network conditions. Figure~\ref{fig:architecture} depicts the architecture of AR devices. The traffic of each flow group is first load balanced over available access networks ($e.g.,$ internet, MPLS) using a \emph{routing agent} and then a scheduling engine at each port (each access network link), controlled by a \emph{QoS agent}, applies a QoS policy, $i.e.,$ the WFQ based RL approach we describe later on. The monitoring block provides information on the network at path and flow group levels such as jitter, delay, and throughput metrics, some of which are factored into our deep learning decision-making. The focus of this paper is on the smart queuing part of the aforementioned architecture. In what follows, we consider that the routing policy is already decided, and we discuss only traffic scheduling, with the integration of the two being the subject of our future works. In practice, we expect the routing control loop to be much slower than the QoS one, $i.e.,$ decisions by the latter will be taken at a much faster pace. This allows for our current approach of separating the two, wherein the QoS decisions are being taken during what constitutes a steady state for the routing.\\
\indent Our objective is to satisfy SLAs for classified network flows. In particular, we aim at meeting performance targets for each flow group in terms of minimum throughput and maximum end-to-end delay. To do so, we enlist the aid of DRL to continuously optimize queuing parameters. In what follows, we discuss our WFQ approach and the QoS agent's role.\\
\indent Finally, we note that in addition to our SD-WAN use case, we also test our proposal in a more generalized network topology, namely the Abilene topology. We show that our learning model is resilient to different topologies and can adapt to the objectives regardless of the scenario at hand.
\subsection{Adaptive Weighted Fair Queuing and DRL Agents}
\indent 
While strict priority queuing is generally used to prioritize traffic, WFQ can be used to maintain fairness, and its weights can be adjusted so that each flow group of traffic receives a bandwidth proportional to its weight. The latter also impacts the resulting end-to-end delay experienced by the flows. Let $\{ 1, \ldots , K\}$ denote the set of flows. In a WFQ scheduler, each flow achieves an average data rate $R_k$ equal to:
\begin{equation}
R_k = \frac{w_k}{\sum_{i=1}^{K}{w_i}}R,
\end{equation}
where $R$ is the total link capacity, and $w_k$ is the weight associated with flow $k$. As such, the greater the weight of the flow is, the higher its service rate and the lower its local queuing delay are (see latency-rate server model~\cite{stiliadis1998latency}). \\
\indent We assume that every flow in the SD-WAN network can, based on packet priority, be classified intro three groups. These flow groups are in order of importance: gold, silver, and bronze.
Each group has its set of minimum throughput thresholds to be attained: $T_g$, $T_s$, and $T_b$ for gold, silver, and bronze, respectively and an equivalent set of maximum end-to-end delay thresholds to be respected: $d_g$, $d_s$, and $d_b$. The objective of the DRL agents for QoS is to assist in meeting these thresholds by learning how to continuously update the weights (increase or decrease) for each flow group served by the WFQ algorithms. Each agent in this MARL deployment is built on top of a WFQ scheduler. While WFQ is used in our SD-WAN use case, the proposed solution is generic enough to handle any other scheduling architecture. The agents observe the throughput and end-to-end delay values attained by the flow groups, and then make individual decisions on whether to increase or decrease the weights for the flow groups that are served at their corresponding nodes. With the delay and throughput values being influenced by how the packets traverse the entire network, inter-agent communications are expected to be a key feature. 
\section{Graph Convolutional Reinforcement Learning For Multi-agent Systems}\label{dgn}
\indent The objective of the deep learning agents is to continuously adjust the weights, either by increasing or decreasing them, for a weighted fair queuing algorithm managing a set of classified network flows. These agents are situated at ingress nodes across the network, such as the numbered ones in our illustrated scenario in Figure~\ref{fig:nettopology}. In this paper, our main proposal utilizes multi-agent graph convolutional reinforcement learning, or DGN, to manage both how the agents learn and communicate. DGN combines the ideas of graph neural networks and deep reinforcement learning. The agents are embedded in a graph $G=(V,E)$, whose topology is related to the computer network in our scenario. The existence of an edge between two agents in this graph means that they can exchange information. Each node (agent) $i\in \mathcal{N}$, where $\mathcal{N}$ denotes the set of agents, has a set of neighbors $\mathcal{B}_i$ with which it can communicate. This collaboration between agents can be parameterized and is dependent on an adjacency matrix $\mathcal{C}$ that defines which agents are neighbors. Limiting agent communication to neighbors reduces what could be costly interactions, in terms of bandwidth and complexity, while keeping the neighborhood present between agents that are likely to impact each other the most.

\subsection{Multi-agent System,
Replay Buffer and Target Network}
\label{masdetails}

\begin{figure}
	\centering
	\includegraphics[width=0.95\linewidth]{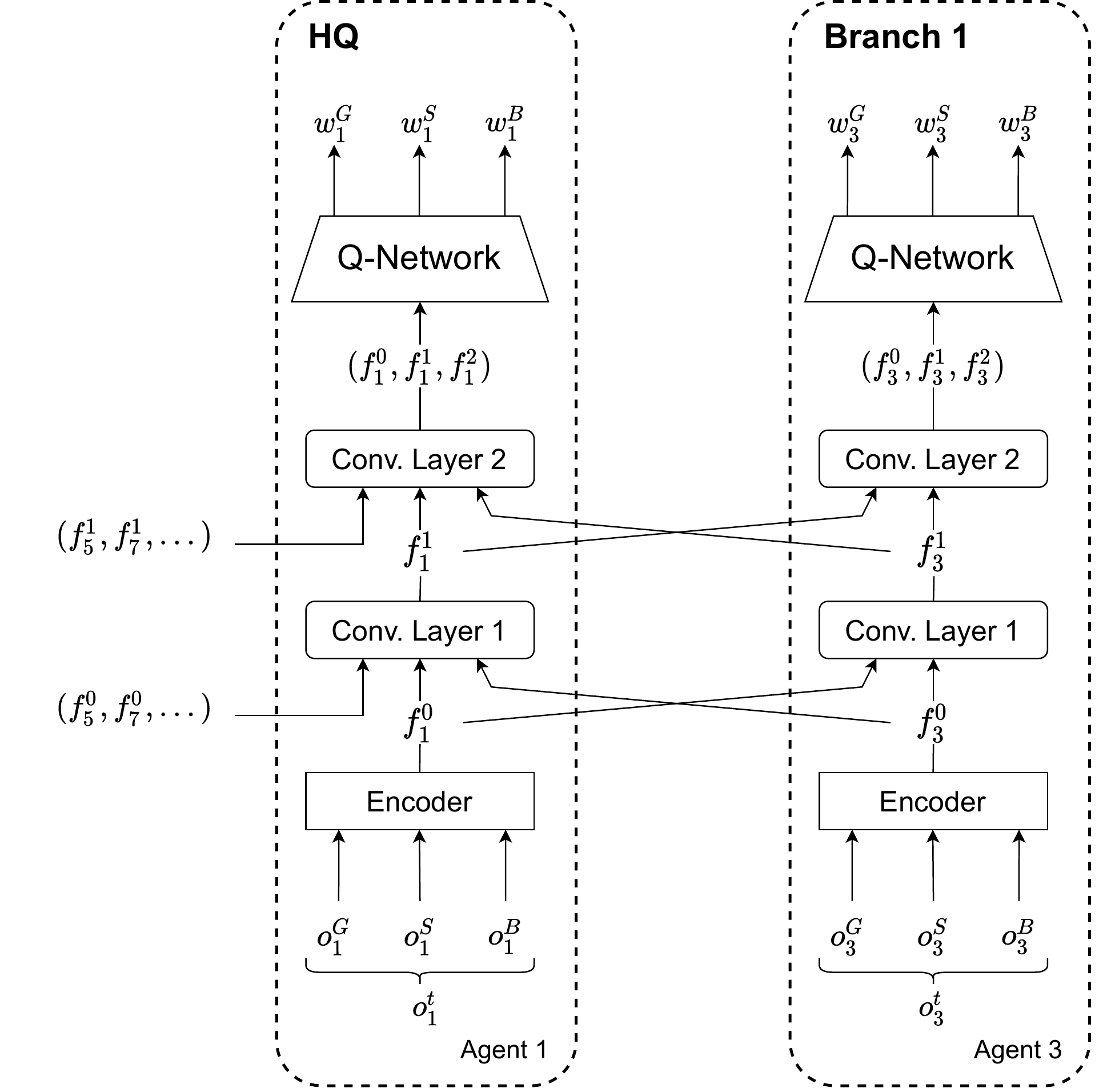}
	\centering
	\caption{Structure of two DGN agents at HQ and Branch $1$}
	\label{dgn1}
\end{figure}
\indent In DGN, the learning problem is formulated as a partially observable Markov decision process. During every time iteration $t$, each agent $i$ receives a local observation from the environment denoted $o^t_i$. The latter consists of a set of values detailing the end-to-end delay and throughput values of the flow groups it is serving. The agent then takes an action $a^t_i$, increasing or decreasing the WFQ weight of each flow group, and as a result is issued a reward $r^t_i$ determined by whether the SLA requirements for the flows groups are met or not. The aim is to maximize the sum of the expected rewards of all the agents. \\
\indent \textbf{Multi-agent collaboration.} Each agent $i$ will run its own reinforcement learning algorithm, whose purpose is to learn how weights $(w^G_i, w^S_i, w^B_i)$, for the gold / silver / bronze flow groups, should change with respect to the local observations and information received from its neighbors. As displayed in Figure~\ref{dgn1}, this reinforcement learning agent is composed of multiple modules. The first module, a multi-layer perceptron (MLP) referred to as an encoder, takes as input the local observations of the agents and extracts the relevant features, referred to as $f_i^0$, of these observations. Once each agent $i$ has its computed features $f_i^0$, it will send them to its neighbors and receive their features, which reflect their own observations.\\ 
\indent In Figure~\ref{dgn1}, agent 1 at the HQ sends $f_{1}^0$ to all the branches it connects to and receives $f_{3}^0$, $f_{5}^0$, etc..., from the respective branches. Agent 1 at the HQ shares information with all the internet branch agents as justified by its adjacency matrix, which details its neighborhood. Recall that in our implementation, agents that share links at the network layer are considered to be neighbors. These features will be the input of the second module, which is a convolutional neural network. Similar to the encoder, the role of the convolutional network is to extract the relevant features of the combination of the local observations and the features received from neighbors. As suggested by the figure, a multiple convolutional layer module can be used. Each layer takes as input the features computed by the preceding convolutional layer, as well as a new set of features received from the neighbors. In our work, we consider two convolutional layers. Similarly, as it is performed by distance vector routing to learn the shortest path by exchanging routing tables with neighbors, the exchange of features between agents will permit the agents to obtain local knowledge from agents that are at a distance $h$ from them, where $h$ denotes the number of layers in the second module. For example, the second convolutional layer of node 3 at branch 1 will receive the feature $f_{1}^1$ from HQ node 1, which contains information received from the rest of the branches. Even if these branches cannot communicate directly, the exchange of features with the HQ nodes will permit them to have a full view of the network information. After several stages of convolutional layers, all the information computed will be gathered into a vector of features. The last module is a Q-learning algorithm. It takes as input the features produced by each layer of the convolutional layer. The reinforcement learning algorithm will run on this third module and the decisions, which maximize the expected reward, on the weights will be made by it. 
\indent \textbf{Attention mechanisms.} The convolutional layers of DGN implement attention mechanisms. Convolutional kernels, widely present in convolutional neural networks (CNNs) and image recognition, enable extracting features from images.  In DGN, these kernels integrate the features in the receptive field in order to extract the latent features. They should be able to learn how to abstract the relationship between agents as to integrate their input features. DGN uses a multi-head dot-product convolutional kernel to calculate the interactions between different agents. A more in depth illustration of how attention works in neural networks can be found in~\cite{vaswani2017attention}.\\
\indent \textbf{Replay buffer.} DGN implements a replay (experience) buffer, $i.e.,$ samples are stored in a memory and afterwards randomly sampled for training. This removes any correlation that might exist among consecutive samples. The experiences are of the type ($\mathcal{O}$,$\mathcal{A}$,$\mathcal{O}'$,$\mathcal{R}$,$\mathcal{C}$), where $\mathcal{O}$ is the set of agent observations \{$o_1,...,o_N$\}, $\mathcal{A}$ is the set of agent actions \{$a_1,...,a_N$\}, and as such $\mathcal{O}'$ is the set of new observations \{$o'_1,...,o'_N$\} as a result of the taken actions. $\mathcal{R}$ is the set of rewards issued to the agents \{$r_1,...,r_N$\}, and finally $\mathcal{C}$ = \{$C_1,...,C_N$\} is the set of adjacency matrices for the agents. The adjacency matrices essentially define the neighborhoods for the agents. $C_i$, $\forall$ $i$, is constructed with dimensions ($\vert\mathcal{B}_i\vert$+1) $\times$ $N$, wherein the upper row is a one-hot representation of the index of the agent $i$, and the $kth$ row, $k$ = 2,...,$\vert\mathcal{B}_i\vert$+ 1, is a one-hot rendition of the index of the ($k$-1)$th$ neighbor. Note that the time notation $t$ is dropped from these expressions for the sake of simplicity.\\
\indent \textbf{Target network.} With enough samples in the replay buffer, we are able to train the agents. The training is done with the aid of target networks~\cite{mnih2015human}. A target network is a copy of the agent's main Q-network. Its parameters however are not trained every iteration, but rather updated slowly or every while. This helps root out any instability in training the main Q-network that could arise from consecutive states being very similar. The replay buffer is randomly sampled for a minibatch of size $S$ on which the agent is trained with the purpose of minimizing the loss: 
\begin{equation}
\mathcal{L}(\theta) = \frac{1}{S} \sum_{S}\frac{1}{N}\sum_{i=1}^{N}(y_i - Q(O_{i,C},a_i;\theta))^2,
\end{equation}
where we recall that $N$ is the total number of agents and that
\begin{equation}
y_i = r_i + \gamma \max_{a'}Q(O_{i,C}',a_i';\theta').
\end{equation}
$O_{i,C}$ $\subseteq$ $\mathcal{O}$ represents the observations of $i$'s neighbors. $Q$ represents the Q-function, $\theta'$ the target network parameters, and $\gamma$ is the discount factor. The latter weighs the impact of future rewards. The gradients of the loss of all the agents are accumulated and used to update the main network parameters. The target network parameters are updated smoothly ($i.e.,$ softly) every iteration following:
\begin{equation}
\theta' = \tau\theta + (1-\tau)\theta',
\end{equation}
where $\tau$ denotes the smoothness of the update. If $\tau$=1, then the update is classified as ``hard" and the parameters of the main network are simply copied onto the target network.\\
\indent Finally, we note that during the training phase, as well as during the execution period, the agents are well aware of their neighborhoods, $i.e.,$ their own adjacency matrices. That is to say that they know with which agents they would need to communicate. During this communication, agents share copies of their feature vectors, as illustrated in Figure~\ref{dgn1}. The significance of this overhead is discussed in the results section.
\subsection{DGN based Smart-Queue Management}\label{dgn2}
\indent In our work, we enlist DGN to help with our problematic: to meet stringent SLA requirements for classified network flows. 
The different components presented in DGN are redefined as follows for our problem:
\begin{itemize}
	\item The local observation, in our case, is a tuple representing the end-to-end throughput and delay values attained the flows served by the agent and denoted $\{\overline{T_g}, \overline{d_g}, \overline{T_s}, \overline{d_s}, \overline{T_b}, \overline{d_b}\}$, where $\overline{T_g}$ represents the throughput of the gold flows, $\overline{d_g}$ the average end-to-end delay of the gold flows, and so on. The end-to-end delays are typically measured using in-band network telemetry. 
	
	\item The actions taken by every agent throughout the learning problem consist of either increasing or decreasing the weight of every flow group it is serving (gold, silver, bronze) by a preset constant value $\delta$. Each agent will act on the weights of all three groups simultaneously ($\pm \delta$). This means that, in total, each agent has eight possible actions to take at every iteration.
	\item The reward issued for each agent after it takes an action is relative to whether it has helped meet the requirements for each flow group. Let $\eta_j$ be the reward for meeting required throughput values of flow group $j$, and $\phi_j$ the reward for meeting the delay requirement of the group $j$. For the reward we are aiming to meet the an average end-to-end delay maximum for the flows of the groups. The total reward $r_t$ issued for an action is then computed as follows:
	\begin{equation}
	\omega^{th}_g\cdot\eta_g + \omega^d_g\cdot\phi_g + \omega^{th}_s\cdot\eta_s + \omega^d_s\cdot\phi_s + \omega^{th}_b\cdot \eta_b + \omega^d_b\cdot\phi_b, \label{reward}
	\end{equation}
	where $\omega^{th}_j$ is set to -1 if the required throughput for flow $j$ is not met and +1 otherwise. $\omega^d_j$ is its delay equivalent in regard to meeting the target delay values. Consequently, the agent reward can be negative, $i.e.,$ a penalty.
\end{itemize}
The rewards/penalties for meeting the gold flow requirements are set higher than that for the silver, and for this latter higher than the bronze. That is to say, the agent is better rewarded, alternatively penalized more, for meeting or violating the gold flow requirements than they are for those of the silver and bronze flows, respectively. Note that we can weight the rewards/penalties for the delay with respect to those of the throughput. With gold group flows, for example, $\phi_g$ = $\kappa_g$$\cdot$$\eta_g$. If $\kappa_g$ $<$ 1, the agents are incentivized to meet the throughput requirements ahead of the delay ones, for the gold flows.\\
\indent Note that the throughput and delay are continuous values. Since we cannot learn over a space of infinite states, it is important to discretize it. 
We first need to define the size of the observation space. In our work, we set it to 20. 
This means that for every observed element, we have 20 possible values. Since our observation is made up of six different inputs, the discrete observation space size is a sextuplet, with each element belonging to a set of 20 different values. The window size is computed as the maximum attainable value minus the corresponding minimum for each element of the state tuple. The discrete state is the integer value resulting from subtracting the minimum observation space from the state and dividing the result by the discrete window size.
\begin{algorithm}
	\DontPrintSemicolon
	Define the discrete observation space size: $DISCRETE\_OS\_SIZE = [20, 20, 20, 20, 20, 20]$\\
	Compute the window size: $discrete\_os\_win\_size = (observation\_space.high - observation\_space.low)/DISCRETE\_OS\_SIZE$\\
 	
	\SetKwFunction{FMain}{$get\_discrete\_state$}
	\SetKwProg{Fn}{Function}{:}{}
	\Fn{\FMain{$state$}}{
		int $discrete\_state = (state - observation\_space.low)/discrete\_os\_win\_size$\;
		\KwRet $discrete\_state$\;
	}
	\caption{Discretization of the States}
	\label{discrete}
\end{algorithm}\\
An increased state space would mean the algorithm has more room to explore for better solutions, but would incur a time penalty for convergence. That is to say if we set the discrete observation size to 100 for example, we would be encompassing a lot of additional states/observations. Nonetheless, the space would be too large. Striking the correct balance is mostly a matter of approximation and experimenting.\\ 
\indent Finally, in Algorithm~\ref{dgnalg} we summarize how the training for DGN is done. We detail the process starting from the filling of the experience replay buffer to the training of each agent, up until the update of the target network. The exploration rate $\epsilon$ determines how often the agents explore during learning, and it is kept constant during training in our approach. Convergence can be inferred from the training loss. When the experience buffer has enough samples, the training phase can begin. We train on positive rewards and the terminal state is when maximum reward is achieved. The variable $done$ indicates that the terminal state has been reached.

\begin{algorithm}[!h]
    \small
	\DontPrintSemicolon
	Initialize randomly the main Q network and its target\\
	Initialize the agents and the environment at random states\\
	\While{not converged} 
	{\tcc{Sample phase in the replay buffer}
	
		\For{every agent $i$ $\in$ $\mathcal{N}$}
		{
			Generate a random number $e$\;
			\If{e $<$ $\epsilon$}
			{Choose a random action $a_i$ \;}
			\Else
			{Query the Q-network for the best action based on observation $o_i$\;}
			Agent $i$ gets reward $r_i$ and next observation $o_i'$ \;
		}	
		Store tuple ($\mathcal{O}$,$\mathcal{A}$,$\mathcal{O}'$,$\mathcal{R}$,$\mathcal{C}$,$done$) in replay memory $D$\;
		\tcc{$done_i$ indicates if agent $i$ reached its set target or not}
		
		\If{enough experiences in D}
		{
			\tcc{Training phase}
			Sample a random $minibatch$ of transitions from $D$\;
			\For{every ($\mathcal{O}$,$\mathcal{A}$,$\mathcal{O}'$,$\mathcal{R}$,$\mathcal{C}$,$done$)}
			{
				\For{every agent $i$ $\in$ $\mathcal{N}$}
				{
					\If{$done_i$}
					{
						$y_i$ = $r_i$
					}
					\Else
					{
						$	y_i = r_i + \gamma \max_{a'}Q(O_{i,C}',a_i';\theta').$
					}
				}
				Calculate the Loss $\mathcal{L}(\theta) = \frac{1}{S} \sum_{S}\frac{1}{N}\sum_{i=1}^{N}(y_i - Q(O_{i,C},a_i;\theta))^2,$\;
				Update Q by minimizing the loss $\mathcal{L}$ \;
				Update the target network softly using Q's weights: $\theta' = \tau\theta + (1-\tau)\theta'$\;
				
			}
		}
	}
	\caption{Training the DGN Agents}
	\label{dgnalg}
\end{algorithm}

\section{Deep Q-Learning Approach}\label{dqn}
\begin{figure}
	\centering
	\includegraphics[width=\linewidth]{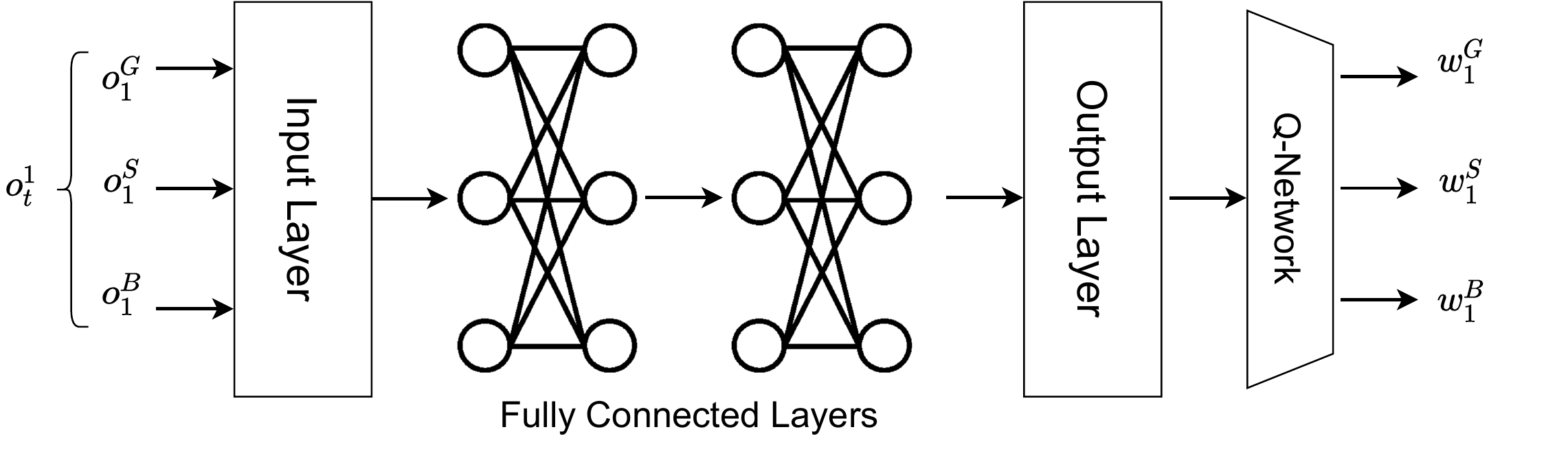}
	\centering
	\caption{Structure of the DQN agents}
	\label{dqn1}
\end{figure}

\indent We present two benchmark multi-agent solutions, based on DQN, to the problem. While DGN is semi-distributed, one of these DQN approaches is completely centralized, and the other fully distributed without any inter-agent communications. Deep Q-learning revolves around the idea of attaching a deep neural network to the traditional Q-learning problem. It aims at solving its memory problem by removing the Q-table and using neural networks to determine the best actions.\\
\indent For the distributed approach, the agents are considered to be fully independent. They each have their own set of states, actions, and rewards, and they each view the environment from their local perspective. No inter-agent communications exist. Figure~\ref{dqn1} shows the structure of the distributed DQN agents in our work. Unlike DGN, wherein we have an encoder and convolutional layers managing inter-agent relations, here we have two fully-connected layers in between the input and output layers. There are no built-in cooperation mechanisms. The actions taken by the agents and the rewards they are issued remain unchanged from before.
In DQN, we again utilize two principle deep learning mechanisms: target networks and replay buffers. The replay buffer is filled with experiences of the type $(\mathcal{O}, \mathcal{A}, R, \mathcal{O}', done)$, $i.e.,$ current observation, action taken, reward received, and the new observation. As in the case of DGN, the variable $done$ indicates if the learning reached its objective or not. The target network, with parameters $\theta'$, is a copy of the main Q-network and is used to stabilize the training. The predicted Q-values of the target Q-network are utilized to backpropagate through and train the main Q-network. However, they themselves are not trained but regularly updated with the values of the main Q-network. In this case, we use a hard update with $\theta' = \theta $.\\
\indent We train on positive rewards, $i.e.,$ the experiences in which the agent does not reach a terminal state are not used to minimize the loss. Once enough experiences are stored in the buffer, the training process starts on randomly selected $S$ minibatches. For every experience, each agent acts as follows:
\[ \begin{cases} 
y_i = r_i, & \text{If $done$ = True} \\
y_i = r_i + \gamma\cdot\underset{a' \in  \mathcal{A}}{max} \  Q(o'_i,a_i';\theta '), & \text{Otherwise}
\end{cases}
\]
The loss, which is minimized using stochastic gradient descent, is then computed as follows:
\begin{equation}
\mathcal{L} = \frac{1}{S}\sum_{S}(Q(o_i,a_i;\theta)-y_i)^2. 
\end{equation}
\indent As for the approach to the weight selection problem itself, things remain virtually unchanged from DGN. We have the same states of states and observations, the same possible agent actions and the reward is calculated in the same manner.\\
For the centralized approach, the agents are trained as if they are one central unit, they interact jointly with the environment, $i.e.,$ they share the same state and observations. They also take their actions jointly, as if it is one superimposed action, and they receive a single reward. The agents are practically sharing a complete vision of the environment and their individual interactions with it. 
\section{Simulation and Results}\label{results}

\indent We now evaluate the proposed multi-agent architecture for smart queuing using ns-3~\cite{ns3} with the deep learning agents being built using Python and TensorFlow.
We simulate SD-WAN network illustrated in Figure~\ref{fig:nettopology}. 
As we considered that routing is controlled by a slower control loop, and it is in steady state (see Section~\ref{archi}), the adjustment of queuing parameters inside transport networks ($i.e.$, internet, MPLS) can be considered independently. For this reason, we only simulated one type of transport network at a time, with a scenario of UDP traffic over the internet, and another of TCP traffic over MPLS, being considered. We simulate HQ-branch links with propagation delays of $10$~ms with capacities of $10$~Mbps each. Small rates are chosen to speed up simulation duration, however we verified the results are not impacted when the bandwidths are of higher magnitude. We used on-off applications for traffic generation and build the WFQ approach over active queue management techniques, namely random early detection (RED). At each branch, origin-destination (OD) flows for two flow groups are generated towards HQ. For instance, at branch 1 we have gold and silver, at branch 2 we have silver and bronze, and at branch 3 we have gold and bronze, etc. The purpose of this variation is to avoid having homogeneity across the traffic treated by the different agents, and thus create a need for agent collaboration.  


\indent \textbf{Agent communications.} In the case of DGN, we recall that the adjacency of the agents in our scenario, $i.e.,$ the neighbors with which each agent can communicate, is defined through the presence of links. In Figure~1, HQ agent 2 communicates with nodes 4, 6, etc..., but not 3 or 5, and so on. In the case of distributed MADQN, no inter-agent communications exist whatsoever. In the case of centralized MADQN, the agents act as if they are one unit, sharing the same environment, states, actions and rewards.

\indent \textbf{Traffic scenario.} Table~\ref{simpara1} details the simulation parameters. The transmit rate of the sources follows diurnal and sinusoidal patterns between 0 and 20~Mbps. The HQ-branch links (ex 1-3, 2-4, etc.) have limited bandwidths and are the links where congestion is likely to occur and impact the general performance of the network. The simulations are done as a series of 300 snapshots, the duration of each being 10~seconds. The duration is enough to achieve a steady state for TCP in our topology, and it has no impact on the eventual results. The weights for the agents are randomly assigned at the start, with that of the gold flow being higher than the silver and the bronze, respectively. The delay metric considered is the average end-to-end delay for flow groups. The simulation parameters can be seen in Table~\ref{simpara1}.\\
\indent \textbf{Benchmarks.} In terms of deep reinforcement learning, we compare our proposal to two multi-agent DQN approaches. One is fully distributed with no agent communications, and the other completely centralized with shared states, actions, and rewards. In terms of traditional approaches to QoS management in queuing, we simulate a classic priority queuing (PQ) algorithm. The latter serves the  packets in descending order of priority. This means that all gold packets are dequeued first, the silver second and the bronze last.\\
\indent \textbf{Training of agents.} Tables~\ref{simpara2} and ~\ref{simpara3} detail the parameters for the MADQN and DGN agents, respectively. When choosing this set of parameters, the objective is to create the smallest neural network capable of addressing the problem. These parameters were set intuitively following models in the state-of-the-art. The exploration rate $\epsilon$ dictates how often during training we take random actions, and how often we utilize the trained model.

\begin{table}[!h]
	\centering
	\caption{Parameters for the simulations} 
	\label{simpara1}
	\begin{tabular}{@{}ll@{}}
		\toprule
		\textbf{Parameter}            & \textbf{Value}                \\ \midrule
		Number of O-D pairs & 10, 4 gold, 3 silver, 3 bronze \\
		Snapshot duration / \# of snapshots              & 10 sec / 300     \\
		$T_g/T_s/T_b$ & 30 / 10 / 5~Mbps \\
		$d_g/d_s/d_b$ for UDP& 0.15 / 0.3 / 0.4~seconds \\
		$d_g/d_s/d_b$ for TCP& 0.1 / 0.15 / 0.2~seconds \\
		Delay to throughput relevance $\kappa_g, \kappa_s, \kappa_b$ & 0.8 \\
		Reward relative to flows G/S/B  & $3x/2x/x$ \\
		WFQ weight update $\delta$ & 0.03\\
		\bottomrule
	\end{tabular}
\end{table}

\begin{table}[!h]
	\centering
	\caption{Parameters for MADQN agents} 
	\label{simpara2}
	\begin{tabular}{@{}ll@{}}
		\toprule
		\textbf{Parameter}            & \textbf{Value}                \\ \midrule
		Activation function        & ReLu   \\
		$N^o$ of fully connected layers        & 2 each with 128 neurons   \\
		Exploration rate $\epsilon$& starts with 1 and decays to 0.001\\
		$\epsilon$ - decay $\epsilon$ & multiplied by 0.99955 per episode\\
		Discount factor $\gamma$             & 0.99                          \\
		Training batch size             & 32            \\
		\bottomrule
	\end{tabular}
\end{table}
\begin{table}[!h]
	\centering
	\caption{Parameters for DGN agents} 
	\label{simpara3}
	\begin{tabular}{@{}ll@{}}
		\toprule
		\textbf{Parameter}            & \textbf{Value}                \\ \midrule
		$N^o$ of convolutional layers       & 2    \\
		$N^o$ of encoder MLP layers&2\\
		$N^o$ of encoder MLP units          & (128,128)                         \\
		Scaling factor $\tau$      & 0.01                      \\
		Discount factor $\gamma$      & 0.99                     \\
		Training batch size             & 32            \\
		\bottomrule
	\end{tabular}
\end{table}
\subsection{Agent Convergence}
\begin{figure}[!h]
	\centering
		\begin{subfigure}{.52\linewidth}
		\centering
		\includegraphics[width=\linewidth]{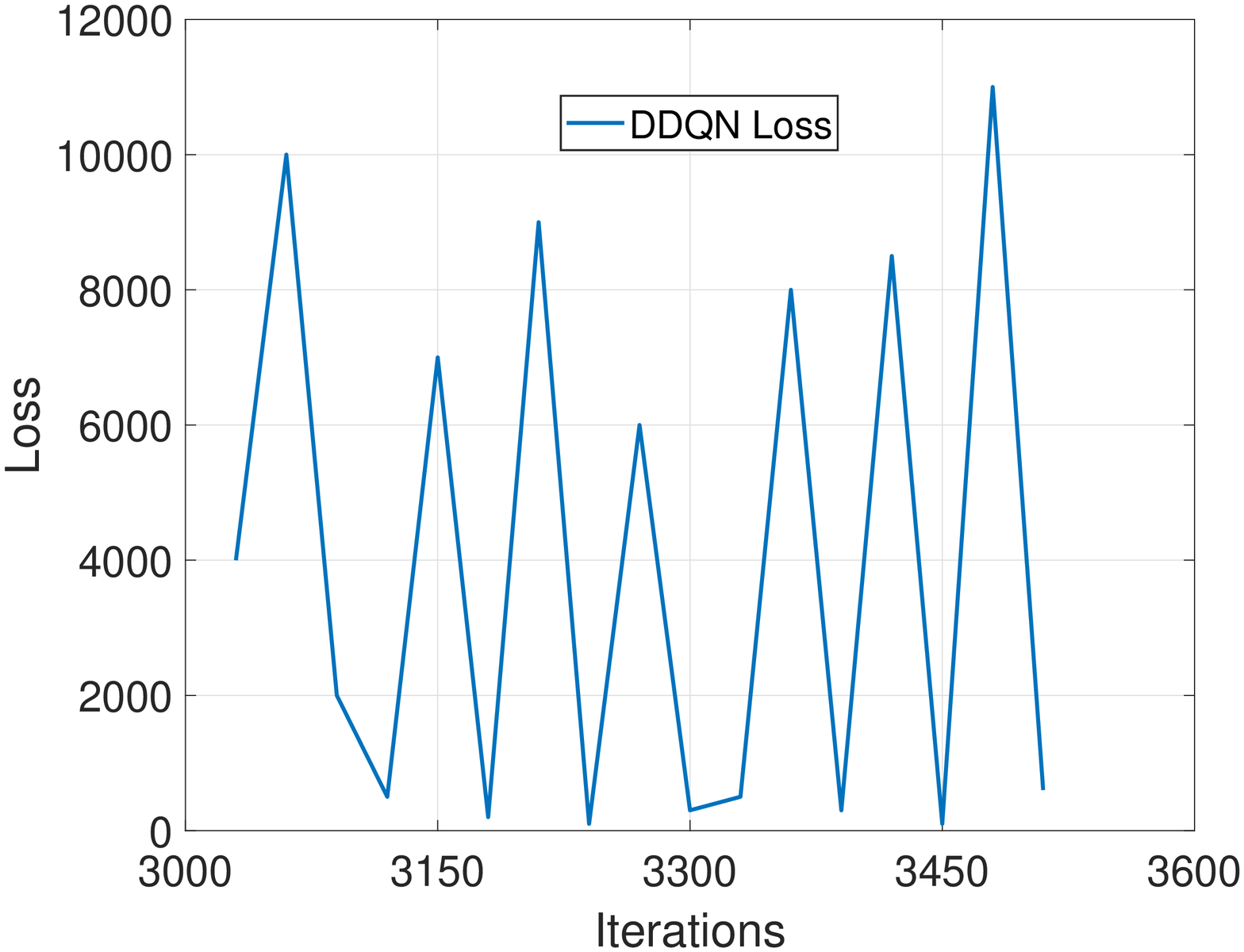}
		\caption{Distributed DQN}
		\label{agent1conv}
	\end{subfigure}%
	\begin{subfigure}{.51\linewidth}
		\centering
		\includegraphics[width=\linewidth]{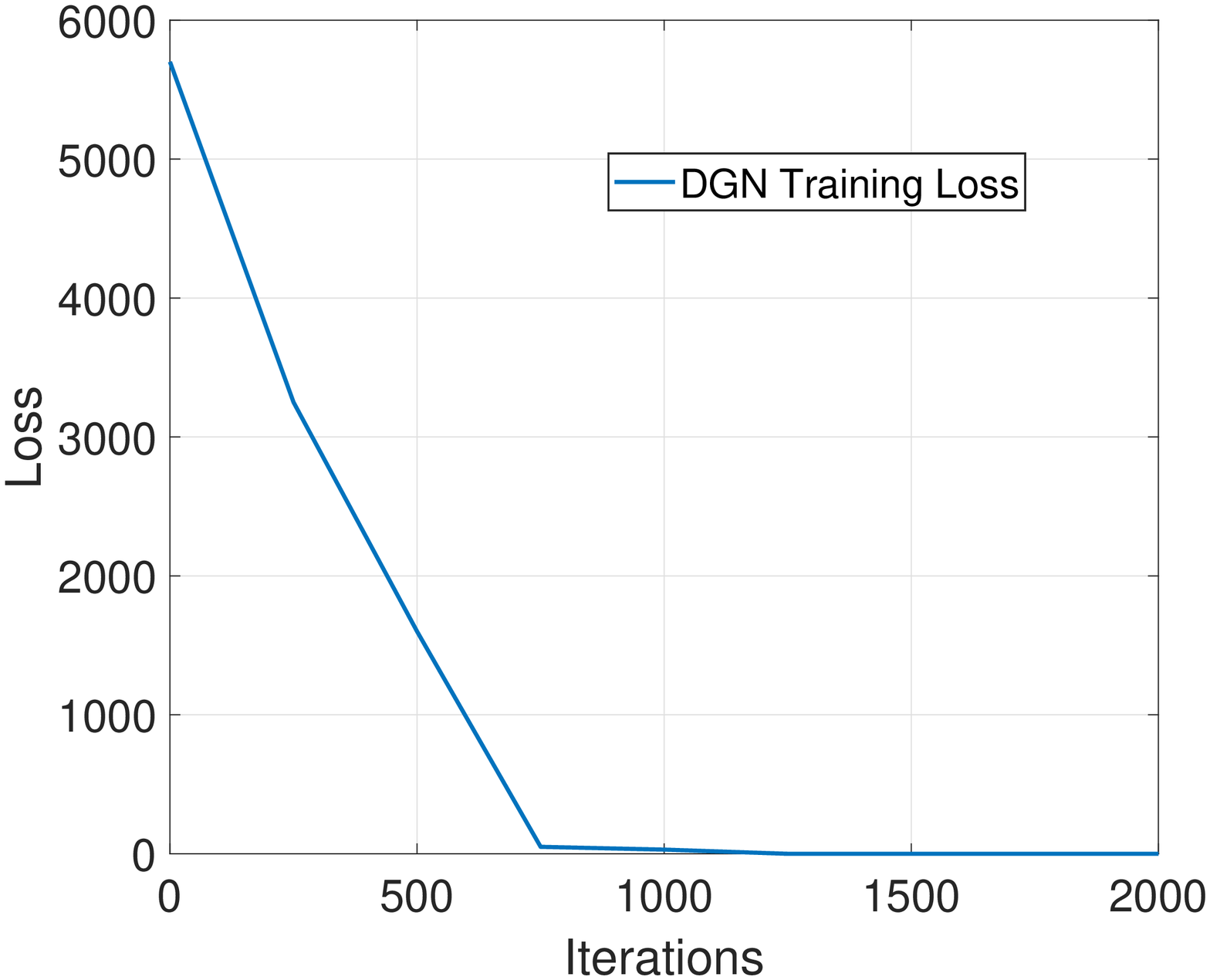}
		\caption{DGN}
		\label{dgnloss2}
	\end{subfigure}
	\caption{Convergence of the learning approaches}
	\label{hetroconv}
\end{figure}

\indent We first assess if the agents converge or not. For the multi-agent distributed DQN approach, Figure~\ref{agent1conv} tracks the loss function for a DQN agent after 3000 training iterations.
 \\
\begin{figure*}
	\begin{multicols}{2}
	\centering 
	\begin{subfigure}{\linewidth}
		\includegraphics[width=\linewidth]{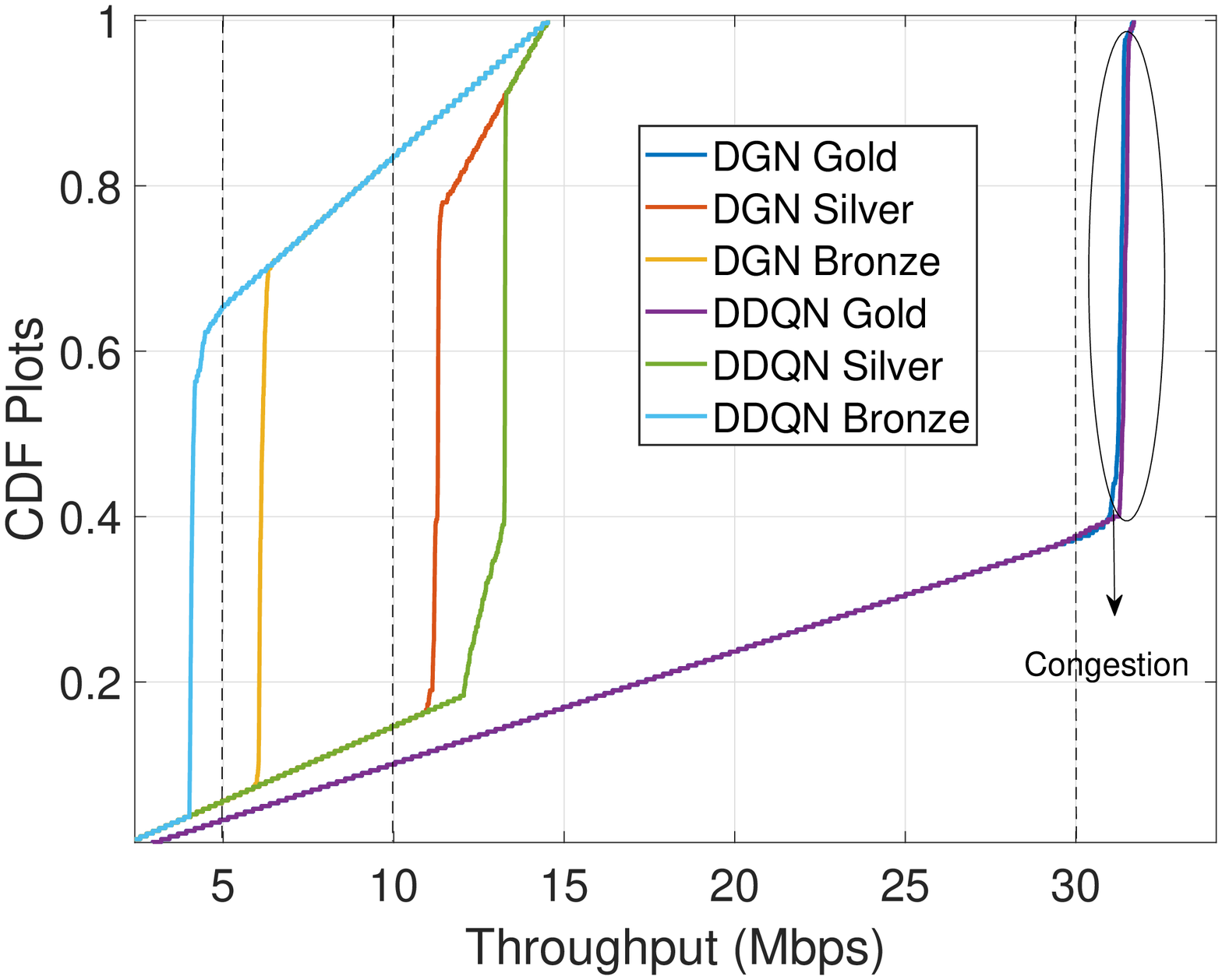}
		\caption{}
		\label{hetrowfqudpDGN}
	\end{subfigure}%
	\begin{subfigure}{\linewidth}
		\includegraphics[width=\linewidth]{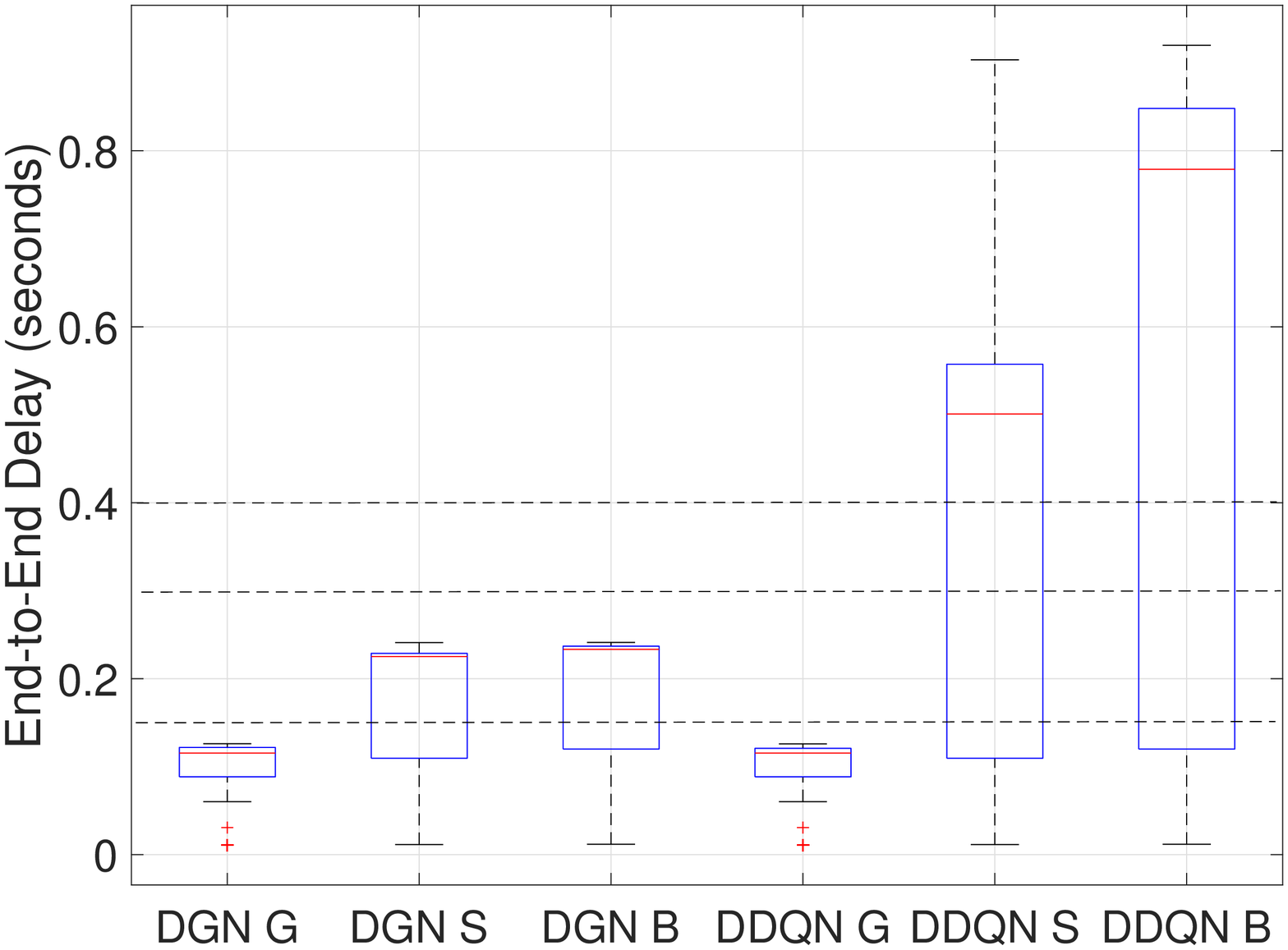}
	\caption{}
   \label{hetrowfqudpDGNDelay}
	\end{subfigure}%
   \end{multicols}
\begin{multicols}{2}
	\begin{subfigure}{\linewidth}
		\includegraphics[width=\linewidth]{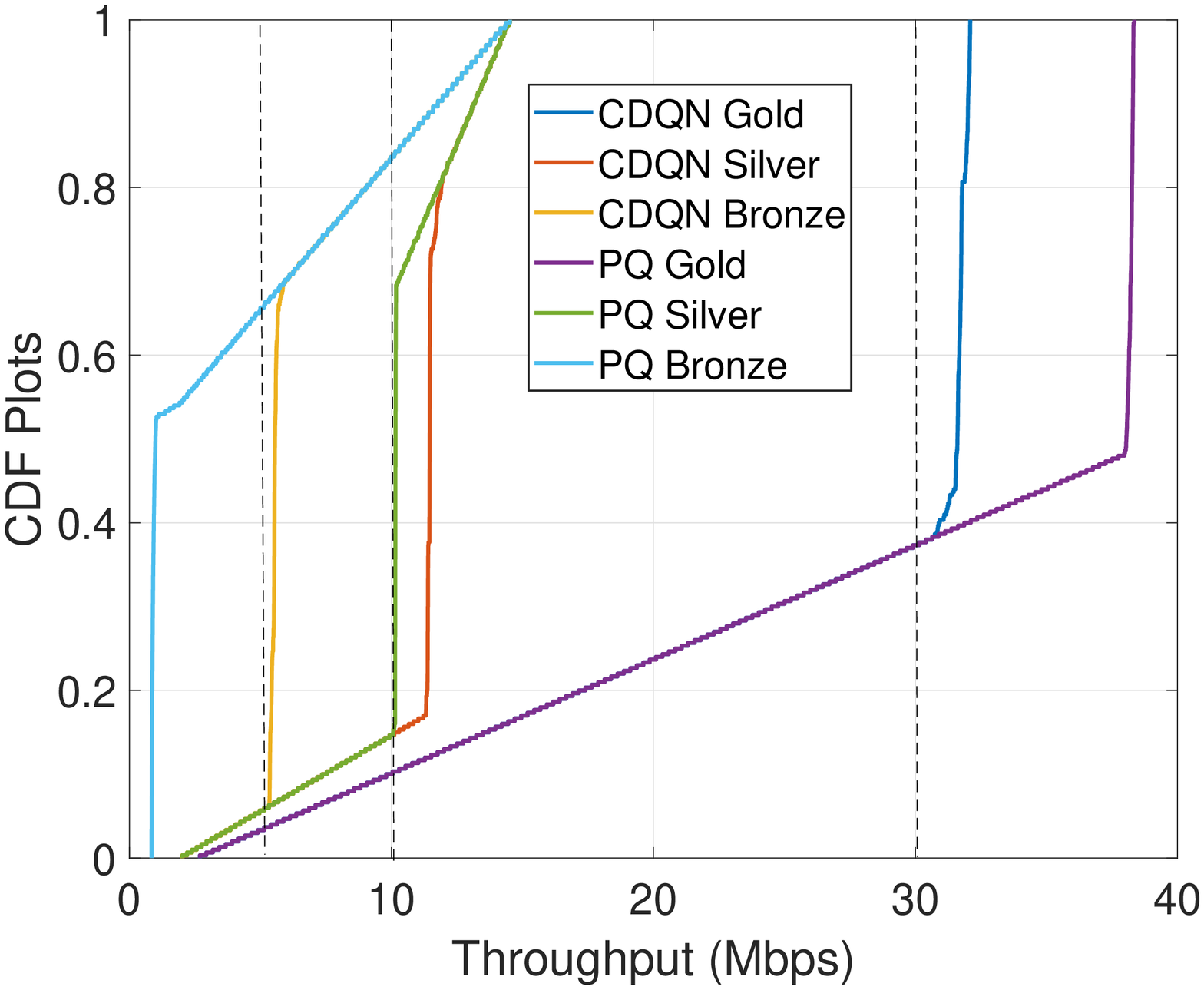}
	\caption{}
    \label{madqnhetrotp1}
    	\end{subfigure}%
	\begin{subfigure}{\linewidth}

		\includegraphics[width=\linewidth]{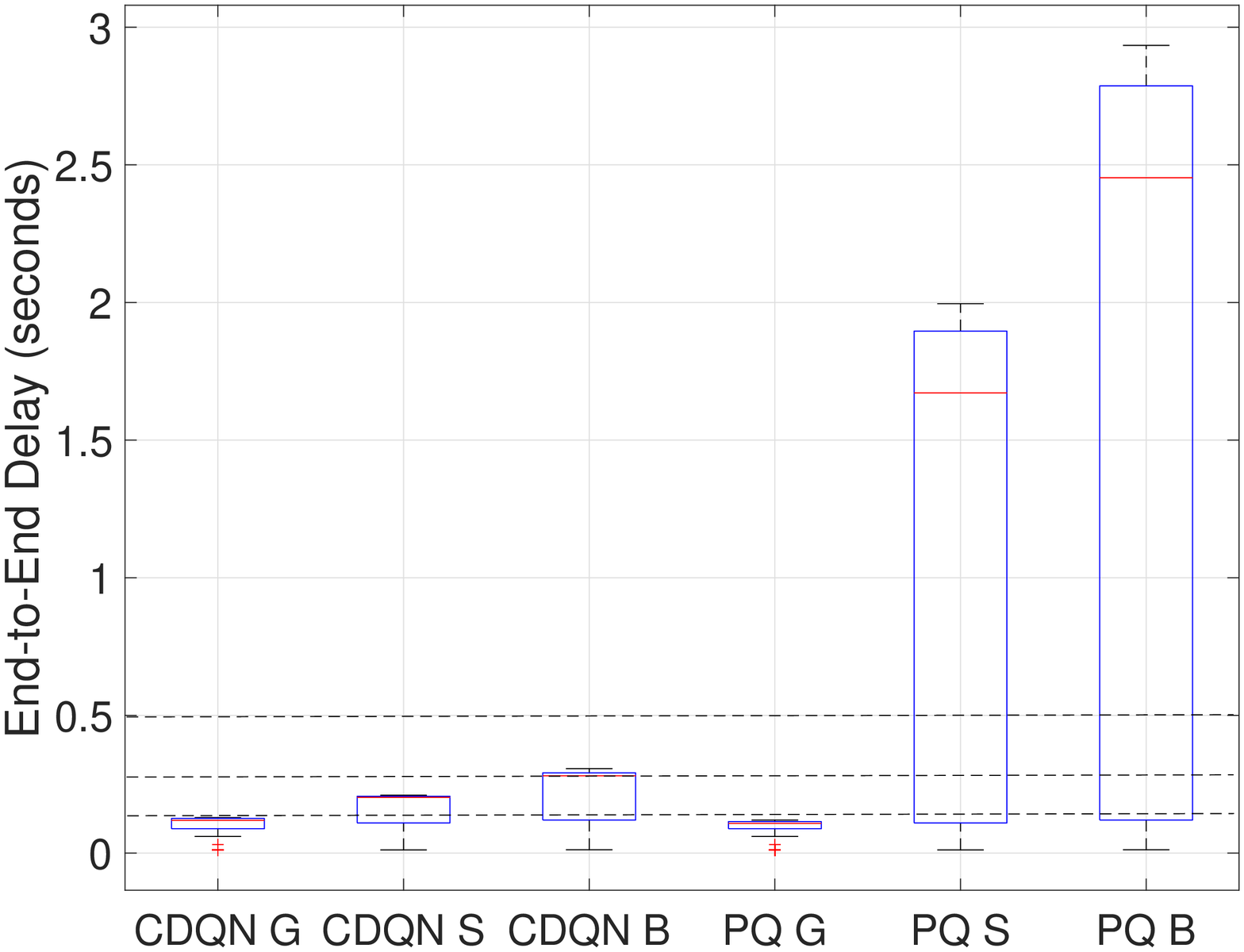}
	\caption{}
    \label{madqnhetrotp1delay}
    	\end{subfigure}%
\end{multicols}

	\caption{UDP traffic: DGN vs. Distributed MADQN (DDQN) (a) and (b), Centralized MADQN (CDQN) vs. PQ in (c) and (d). G: Gold, S: Silver, B: Bronze. SD-WAN scenario.}
	\label{fig:images2}
\end{figure*}
\begin{figure*}
	\begin{multicols}{2}
		\begin{subfigure}{\linewidth}
			\includegraphics[width=0.95\linewidth]{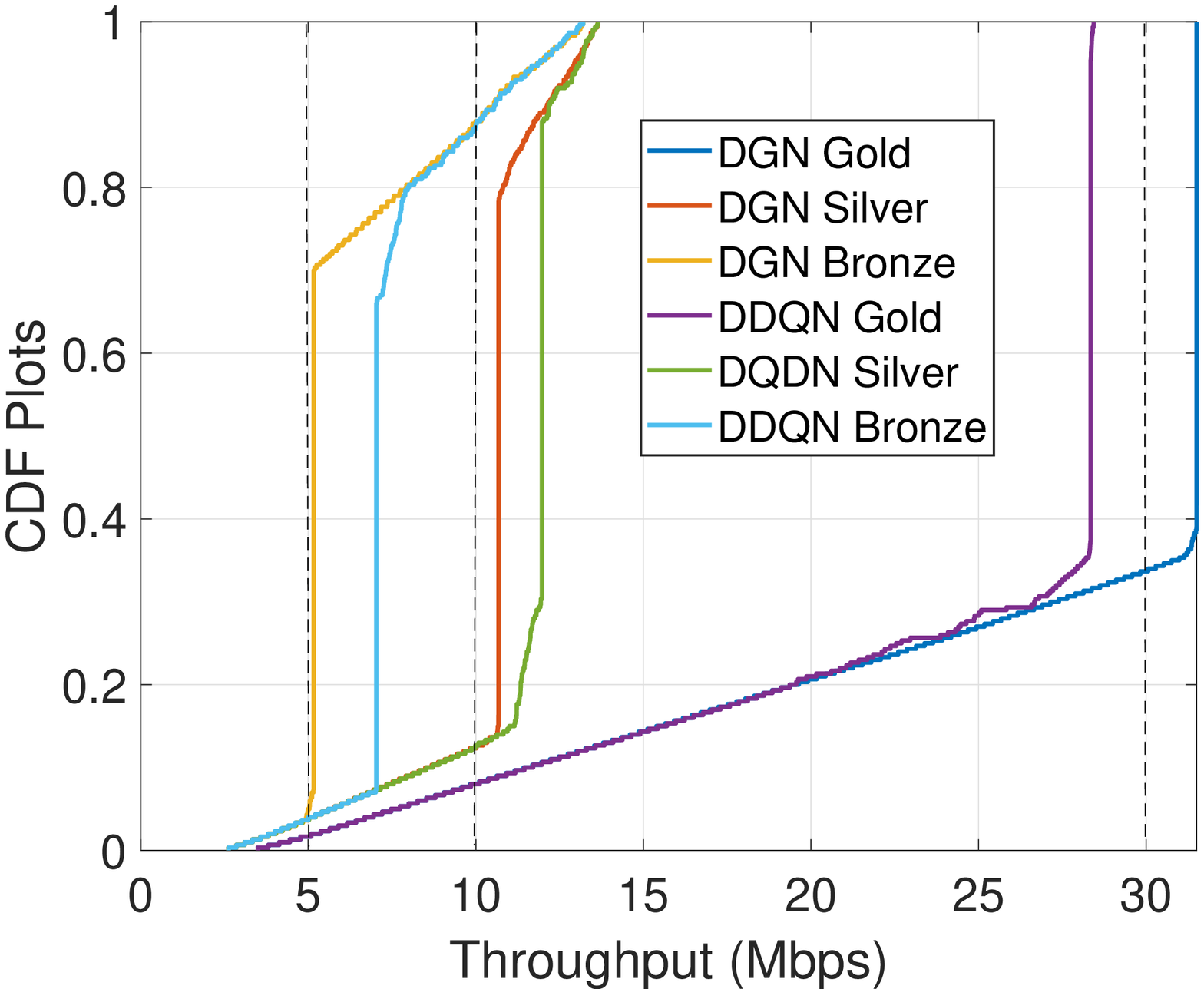}
			
			\caption{}
			\label{hetrogcdqnudp}
		\end{subfigure}
		\begin{subfigure}{\linewidth}
			\includegraphics[width=\linewidth]{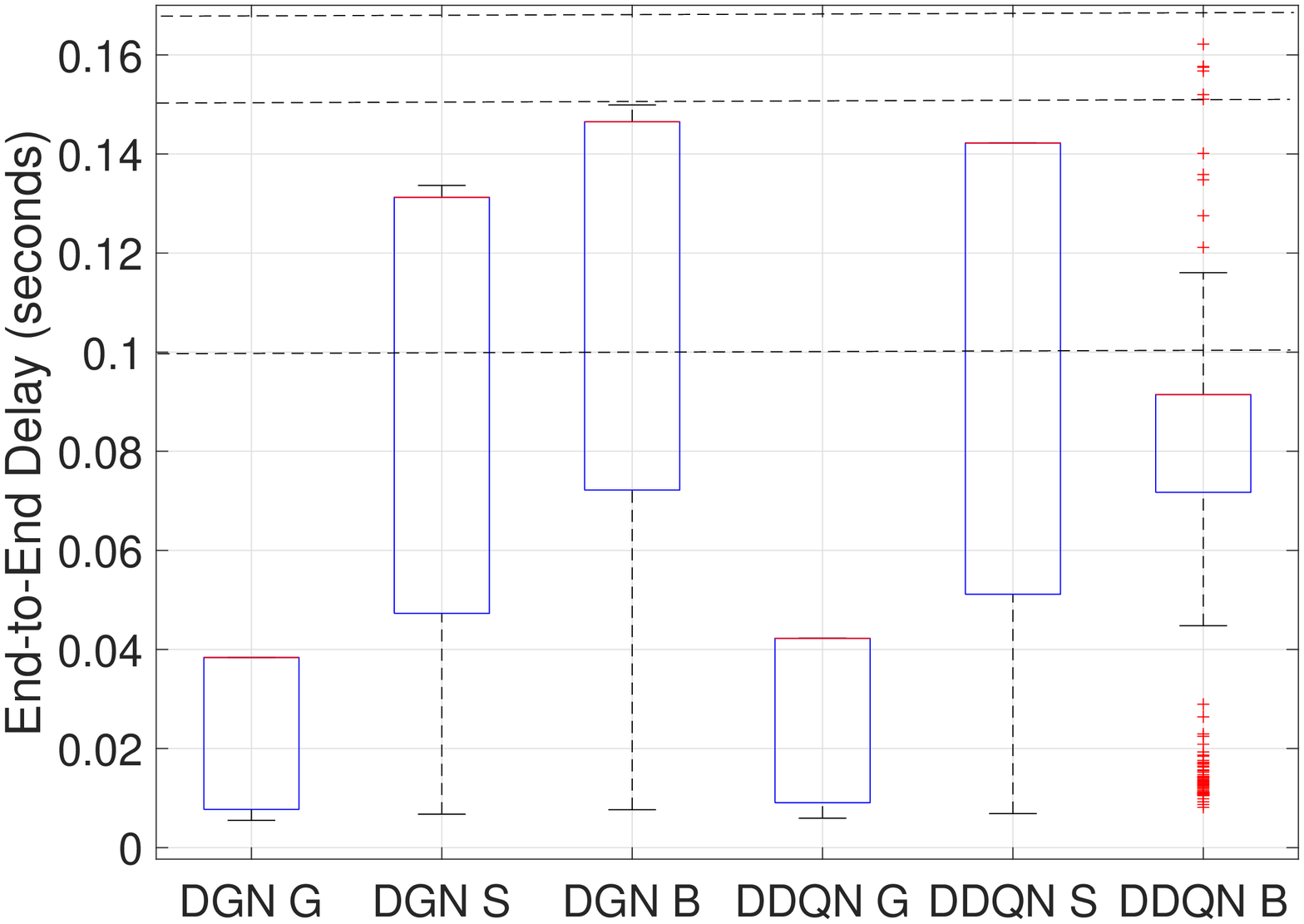}
			\caption{}
			\label{hetrogcdqnudpdelay}
		\end{subfigure}%
	\end{multicols}
	\begin{multicols}{2}
		\begin{subfigure}{\linewidth}
			
			\includegraphics[width=0.95\linewidth]{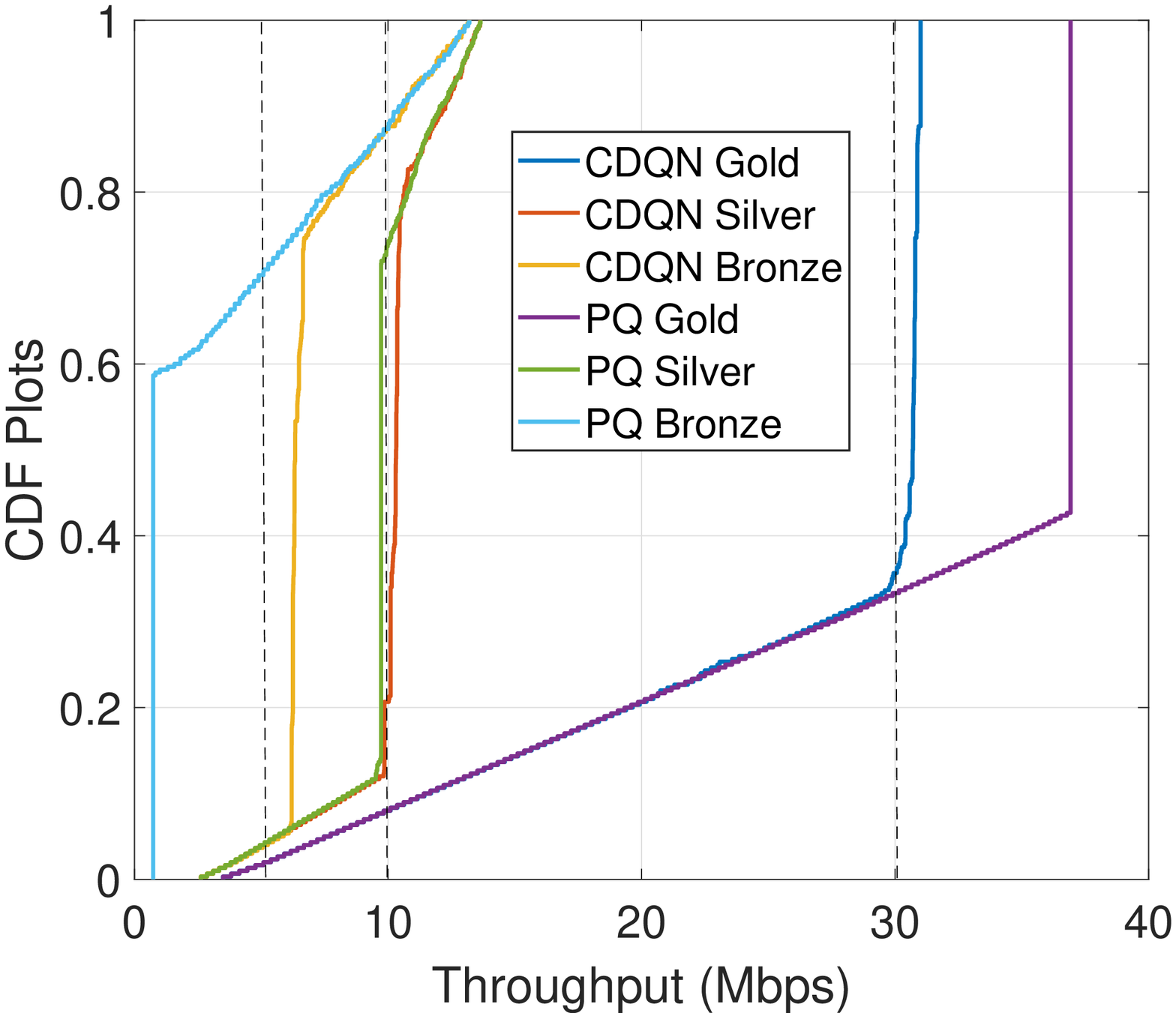}
			\caption{}
			\label{hetrogTCPcmadqn}
		\end{subfigure}%
		\begin{subfigure}{\linewidth}
			
			\includegraphics[width=\linewidth]{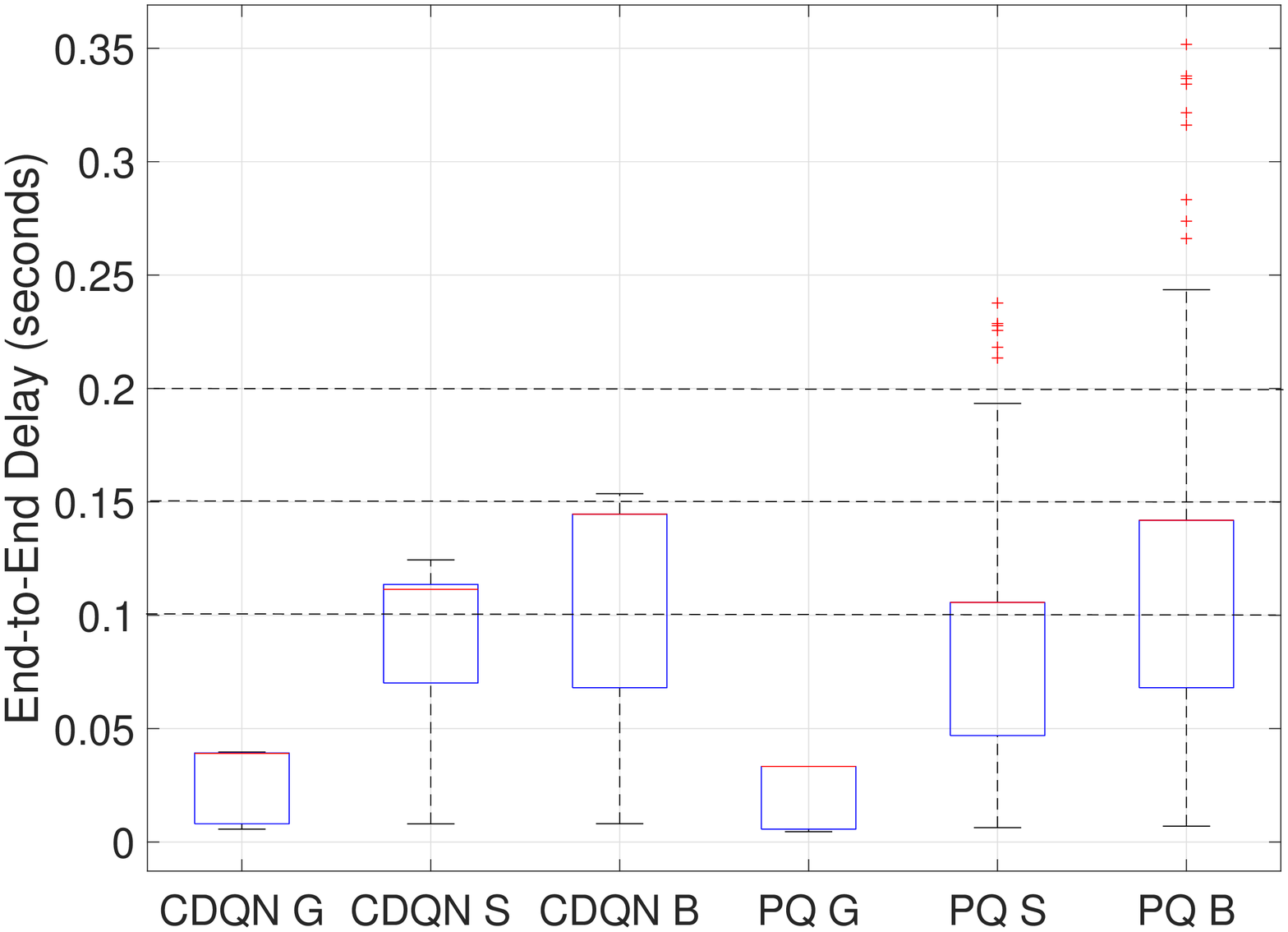}
			\caption{}
			\label{hetrogTCPcmadqndelay}
		\end{subfigure}
	\end{multicols}
	\caption{TCP traffic: DGN vs. Distributed MADQN (DDQN) (a) and (b), Centralized MADQN (CDQN) vs. PQ in (c) and (d). G: Gold, S: Silver, B: Bronze. SD-WAN scenario.}
	\label{fig:images3}
\end{figure*}
 The algorithm is not converging. Even if more time is given for the training, this oscillation remains present. This is mainly down to the completely distributed nature of this distribution. There is a lack of communication between agents in a scenario that requires it, and this is confirmed by the fact that the centralized approach converges without an issue.  For the DGN approach, we track the loss averaged across all the agents. The DGN approach was able to converge. This is illustrated in Figure~\ref{dgnloss2}, where at around 800 training episodes the loss tends towards zero. 
\subsection{UDP Scenario}
\indent First, we consider simulations with UDP traffic. Starting with DGN, we show a cumulative distribution function (CDF) plot with the throughput values attained by the different flow groups throughout all the snapshots. Figure~\ref{hetrowfqudpDGN} has the results. The vertical dashed lines show the SLA requirements in terms of throughput per class. The semi-vertical lines, for each flow group and each DRL solution, in the plots show how the network behaves when the traffic sources are transmitting at a congestion causing rate on the links. It can be seen as a sort of steady state. It is during this period that we are mainly concerned in verifying that the SLAs are met. When there is no congestion, in the lower parts of the diurnal traffic patterns for instance, the traffic sources are transmitting at a rate lower than that needed to maintain the required throughput thresholds. As such, we do not take throughput SLA violations in this region into consideration. We note that the DGN approach can always meet the required throughput. When the network is congested, the throughput for the bronze flow is just above 6~Mbps, for the silver flow about 11~Mbps, and for the gold flow is around 31~Mbps. All above the required throughput values of 30, 10, and 5~Mbps for gold, silver, and bronze flow groups, respectively.\\
\indent We additionally look at how DGN performed in terms of the delay attained by the different flows. Figure~\ref{hetrowfqudpDGNDelay} has box plots with the results. The requirements are 0.15, 0.3, and 0.4~seconds for the gold, silver and bronze flows, respectively. We note that DGN was able to meet all of these thresholds.\\
\indent We observe on the other hand the results for distributed MADQN agents. Figure~\ref{hetrowfqudpDGN} has CDF plots with the results. The latter validate what is seen in the training convergence trend. The demands for the bronze flows are not met, and are below the 5~Mbps mark. While the algorithm does not converge, we did show that the average reward does go into the positives for multiple aggregated iterations as time goes on. As such, the algorithm still manages to meet certain demands.\\
\indent In terms of the delay, the box plots of Figure~\ref{hetrowfqudpDGNDelay} show a similar trend. The delay requirements for the silver flows are violated in more than half the instances. Out of six total constraints, distributed MADQN violates three.\\
\indent Nonetheless, the centralized version of MADQN delivers the required thresholds. In Figure~\ref{madqnhetrotp1}, we note that the throughput values lie around 5.5, 11.4, and at above 31~Mbps for the bronze, silver, and gold flows, respectively. All above the required mark. Figure~\ref{madqnhetrotp1delay} shows that median delay values for the centralized MADQN flows at 0.128, 0.209, and 0.298~seconds meet all requirements. This however is not the case for PQ. We see in the same figures that its bronze flows' throughput is less than 1~Mbps, while the maximum delay values, for both the silver and bronze flows, are around 2 and 3~seconds, respectively. Both are in violation of the required thresholds. Out of the six constraints, PQ meets only three.\\
\indent In conclusion, the communication between agents was key to addressing the problem. The centralized MADQN approach was able to meet the demands unlike its distributed counterpart, highlighting that it is not an issue of the deep learning mechanism being used. DGN provides a solution to the problem without relying on the unrealistic centralized training and execution of the centralized MADQN approach.
\subsection{TCP Scenario}
\indent Similarly, we now look at the results in the case of TCP traffic. Figure~\ref{hetrogcdqnudp} has the throughput results for DGN. Again, DGN meets all the required demands. When the links are congested, the plots show throughput values of about 5.1, 10.6, and 31.5~Mbps for the bronze, silver, and gold flows. All above the set requirements of 5, 10, and 30~Mbps, respectively. The same cannot be said for distributed MADQN. Figure~\ref{hetrogcdqnudp} shows that the throughput requirement for the gold group flows, sitting at around 28~Mbps, was violated.\\
\indent We additionally observe the delay results as reported in the box plots in Figure~\ref{hetrogcdqnudpdelay} for DGN. The required delay thresholds are set at 0.1, 0.15 and 0.2~seconds for the gold, silver, and bronze flows. The DGN agents meet all these requirements. Distributed MADQN meets these demands, but in general with higher mean delay values compared with DGN. Furthermore, the centralized version of MADQN was able to meet all the required throughput and delay thresholds. Figure~\ref{hetrogTCPcmadqn} shows throughput values at around 6.2, 10.1, and 31~Mbps for the bronze, silver, and gold flows, respectively. The same cannot be said regarding priority queuing. The silver flows throughput is about 9.7~Mbps and the bronze about 0.73~Mbps, both in violation of the requirements.\\
\indent In terms of delay, Figure~\ref{hetrogTCPcmadqndelay} shows that centralized MADQN group flows have maximum delays at around 0.04, 0.12, and 0.15~seconds, respectively. All within the required margins. As for PQ, it fails to meet the delay requirements for both the silver and bronze flows, with maximum values recorded at 0.23 and 0.35~seconds, respectively.\\
\indent The results with TCP traffic validate the conclusions of their UDP counterparts. The lack of agent communications in the decentralized MADQN approach caused the algorithm to be inefficient. The results also show that our proposals are much more equipped to deal with the problem than priority queuing. 
\subsection{Impact of Varying the Number of Convolutional Layers}
As discussed in Section~\ref{masdetails}, the number of convolution layers controls the communication between agents. Indeed, we verified experimentally the significance of the number of convolutional layers in the DGN agent modules. To do so, we ran the same experiment from before but this time with the convolutional layer module containing only one layer.
\begin{figure}[!h]
	\centering
	\includegraphics[width=\linewidth]{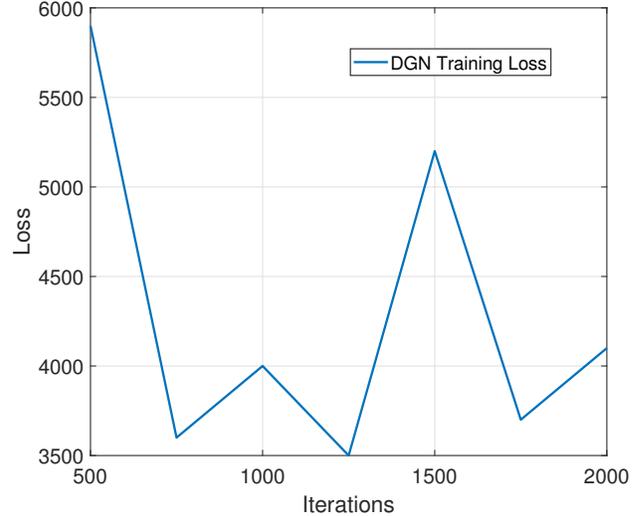}
	\centering
	\caption{DGN convergence with one convolutional layer. SD-WAN scenario.}
	\label{dgnloss2}
\end{figure}\\
The main reason being that agents controlling branches are not able to efficiently collaborate via the HQ. The presence of only one convolutional layer, means that the 2-hop communications needed in this scenario, are not available.
\subsection{On Agent size, Complexity, and Communication overhead}
\indent When setting neural network parameters for the deep learning agents, we always seek the smallest working configuration. That is independent of the MARL agent setting, whether the system is centralized, distributed or semi-distributed. In terms of size of agents, in bytes on disk, DGN agents are considerably larger. DGN has more layers, more structure, and even its own Q-network. A trained DQN agent takes an average 0.267~MBs of space, while DGN ones consume 1.9~MBs.

For DGN, during the agents' execution phase, the agents would need to communicate their feature vectors, $i.e.,$ the output of the convolutional layers. We assess the incurred overhead using two methods. First, in order to quantify the amount of communications involved, as discussed in~\cite{zhang2019efficient}, the overhead is defined as a function the total number of pair of agents that communicate during a certain time instance $t$ $\in$ $T$, denoted $g_t$, and the total number of agent pairs $R$ as:
\begin{equation} 
    \beta = \frac{\sum_{t=1}^T g_t}{RT}
\end{equation}
A ratio closer to one, would mean all the agents are talking with each other. One closer to zero, means agents are barely communicating. In the case of the SD-WAN scenario, the ratio is 0.18 for our DGN approach. This indicates a very small overhead and communications limited to where needed. \\
\indent Additionally, we compute the bandwidth required for such inter-agent communications. In our DGN scenario, the agents refresh their policies every 10 seconds. At that time they need to communicate messages equal to the number of convolutional layers they have. The size of each message is equal to the size of the feature vector. For our implementation, we have 2 convolutional layers. The output of each is 1x128. Assuming a 64 bit machine, we would need 4 bytes to store each of these values. That means on each link between two communicating agents, we would only need 0.8192~kbps of reserved bandwidth for inter-agent communications.\\
\indent We would also note that in this work, we considered a synchronous execution of agents so that they exchange information at the time is it needed by their neighbors. An interesting development would be to consider the asynchronous setting.
\begin{figure*}
	\begin{multicols}{2}
		\begin{subfigure}{\linewidth}
			\includegraphics[width=1\linewidth]{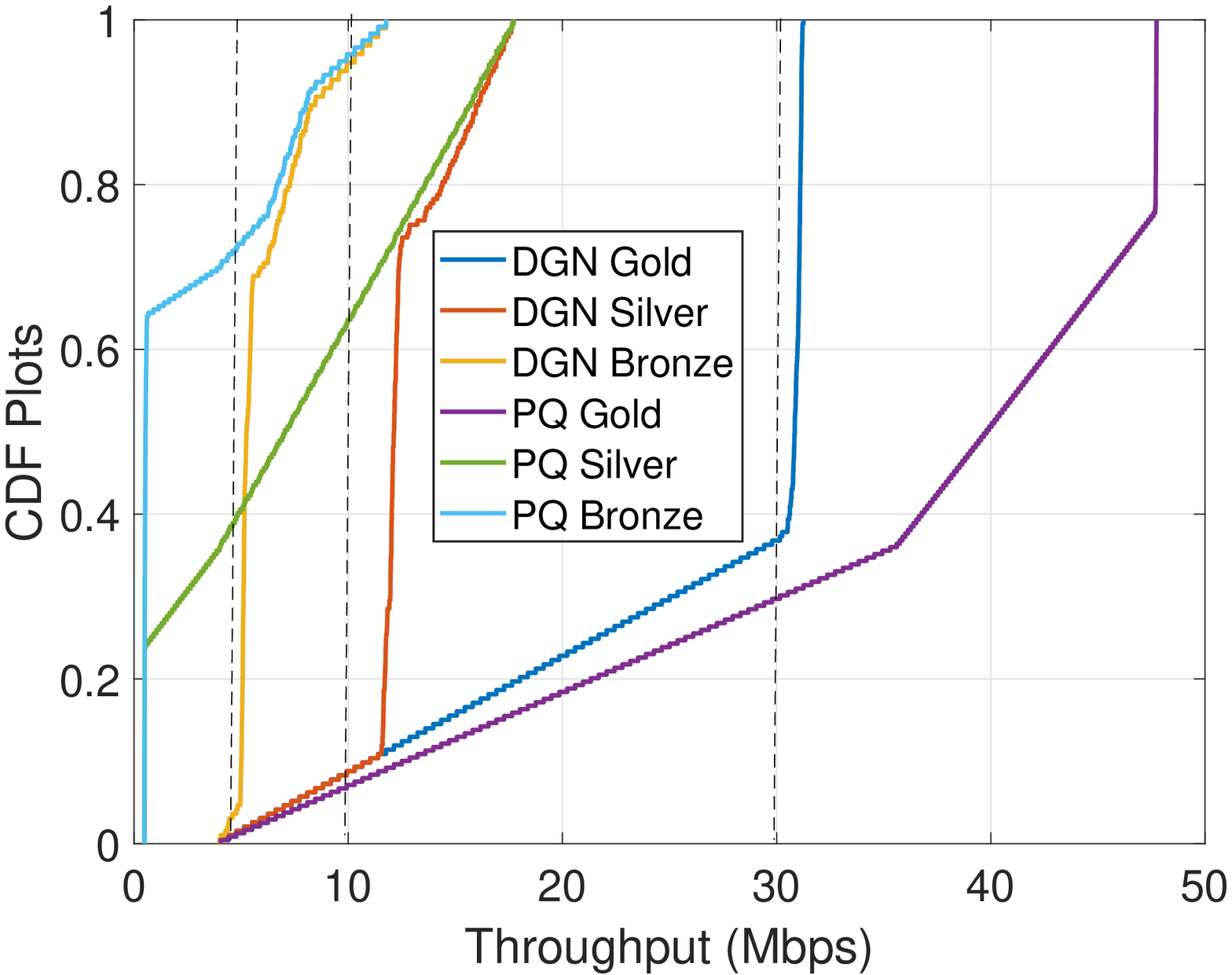}
			
			\caption{}
			\label{albtp}
		\end{subfigure}
		\begin{subfigure}{\linewidth}
			\includegraphics[width=0.94\linewidth]{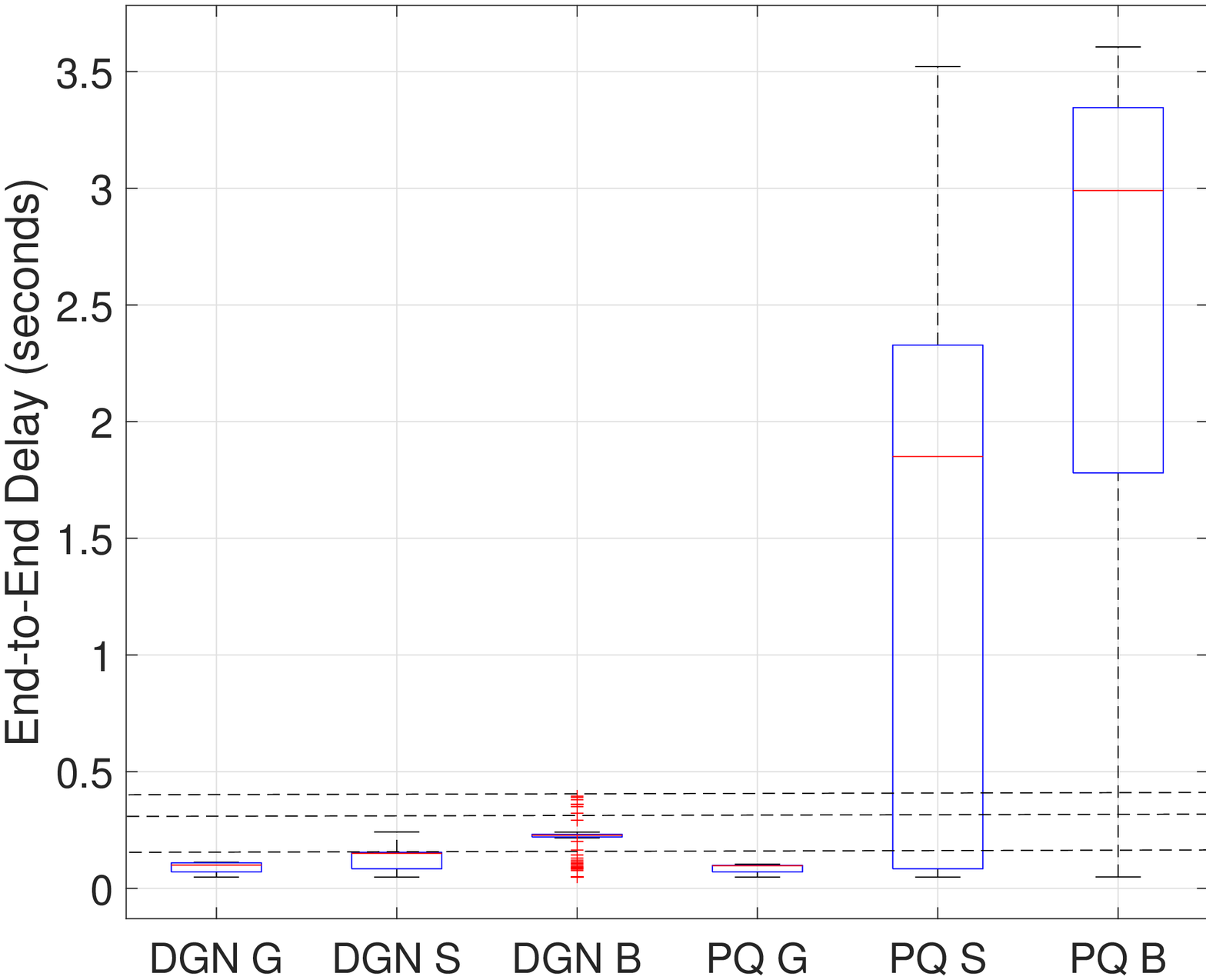}
			\caption{}
			\label{albdelay}
		\end{subfigure}%
	\end{multicols}
	\caption{DGN vs. PQ in terms of throughput (a), and delay (b). G: Gold, S: Silver, B: Bronze. Abilene scenario.}
\end{figure*}
\subsection{Generalized Network Topology}
\begin{figure}[!h]
	\centering
	\includegraphics[width=0.95\linewidth]{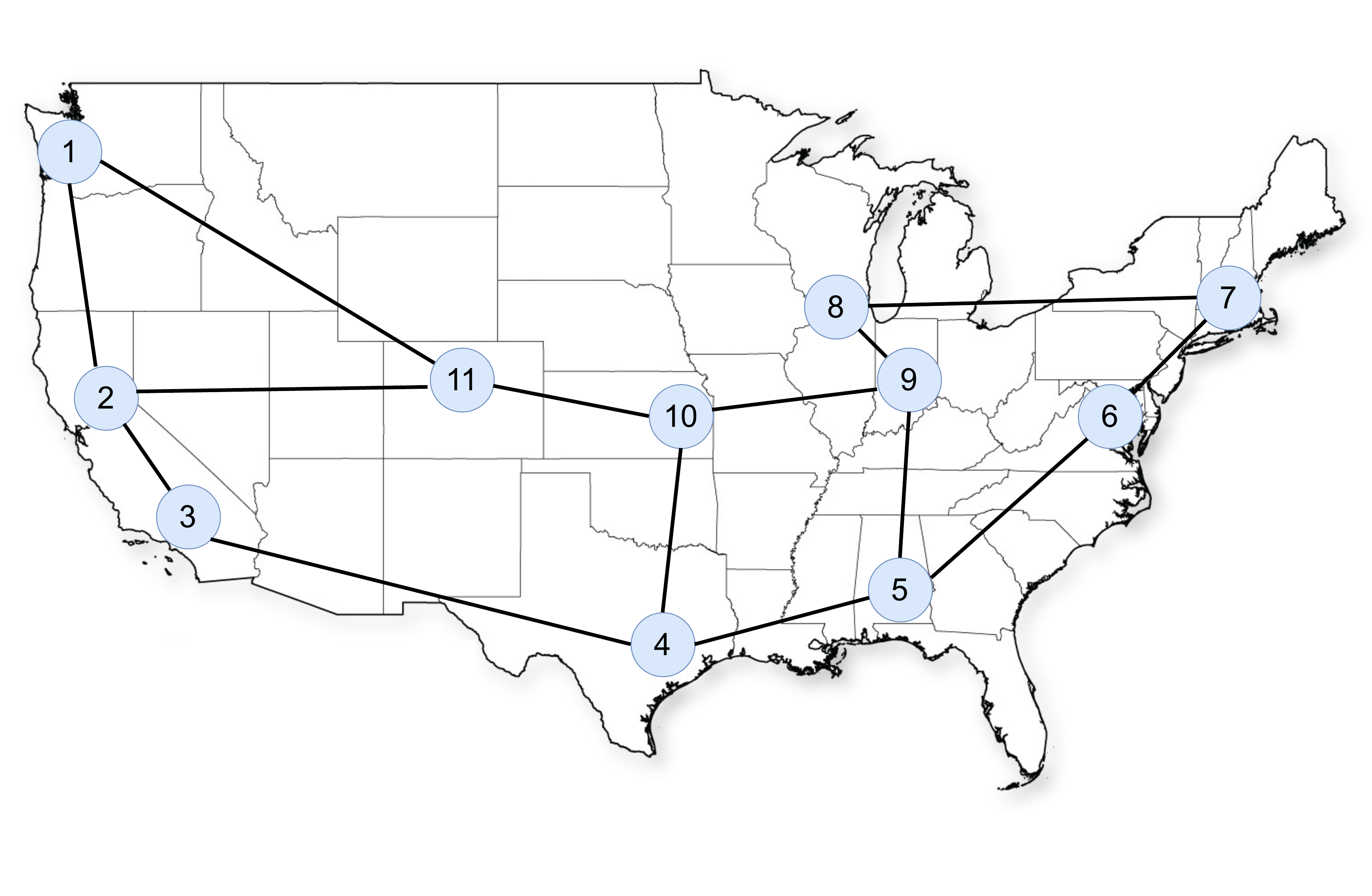}
	\centering
	\caption{Abilene Topology}
	\label{abeline}
\end{figure}

\begin{figure}[!h]
	\centering
	\includegraphics[width=0.97\linewidth]{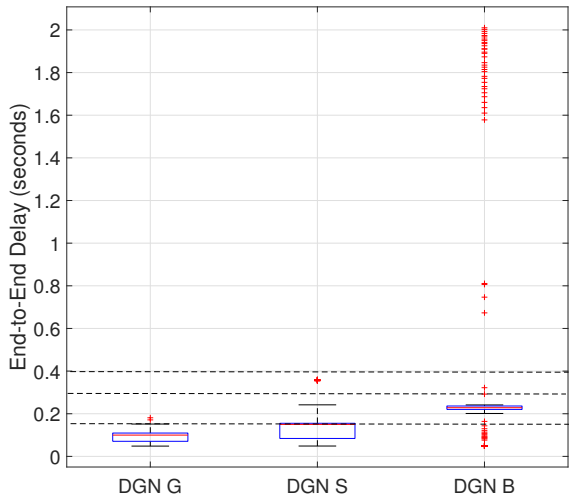}
	\centering
	\caption{End-to-end delay values. One convolutional layer. Abilene scenario.}
	\label{delayab2}
\end{figure}

\indent Finally, we are interested as well in testing our proposal in a more generalized network topology. For that, we choose the Abilene network topology illustrated in Figure~\ref{abeline}. Traffic is generated from hosts connected to nodes 1, 2, and 3, and collected at a destination connected to node 7. Each of these sources generates three flows, one of each type: gold, silver, and bronze. Similar to our previous scenario, we consider that the links interconnecting the main nodes shown in the figure are the ones with bandwidth constraints. Our DGN agents are placed on all the nodes (one through eleven). Two convolutional layers are considered for each agent. At each of the source nodes, we have a gold flow, a silver, and a bronze flow. The sources are of type UDP. Reminder that the source flows are sinusoidal and initially the transmit rates are not enough to meet the throughput requirements. We are concerned with SLA violations only in the period where they can.
\\
\indent Figure~\ref{albtp} compares between our approach and priority queuing (PQ) in terms of throughput. The required thresholds are maintained as before. During congestion on the links, we note that our DGN approach is able to meet all the requirements for all flows with gold flows throughput being between 30.8 and 31.2~Mbps, the silver flows throughput at around 12~Mbps, and bronze flow throughput values just above 5~Mbps. For PQ, the algorithm is capable of meeting the gold flows' requirements, but records violations in both silver and bronze flows' requirements.\\
\indent In Figure~\ref{albdelay}, we compare between the two approaches in terms of the end-to-end delay. The requirements as before are set at 0.15, 0.3, and 0.4~seconds for gold, silver, and bronze, respectively. With DGN, all the flows meet their requirements with maximum values recorded at 0.112, 0.242, and 0.389 for the gold, silver, and bronze flows, respectively. The same cannot be said  for PQ, which shows excessive violations for both the silver and bronze flows.\\
\indent Additionally, we are also interested in measuring the impact of the number of convolutional layers the DGN agents have on their performance. We reduce the number of these layers per agent from two to one, and afterwards repeat the training and the simulations under the same settings. In Figure~\ref{delayab2}, we show the resulting delay values achieved by the DGN agents with the aforementioned structure.
\\
\indent We notice that the approach no longer uniformly meets all the delay requirements, with several infringes recorded for gold and silver flows, and violations in more than 18~\% of the cases for the bronze flows. As with before, the reduced number of convolutional layers causes the DGN to fail in extracting key relations between the different agents, that would have otherwise enabled it to succeed.\\
\indent In conclusion, we showed the resilience of our model and its capability of adapting in different and classic network topologies. We again showed the relevance of the number of convolutional layers when it comes mapping the relationship between different DGN agents. Note that in our current model, we assume that information needed for the agents to make decisions (observations) are available or shared with them. We are currently experimenting with keeping the agents only at ingress and egress nodes, while increasing the number of convolutional layers per agent. This increase will make up for the loss in intelligence from the network center due to the lack of agents. Preliminary results show that we can maintain the SLA requirements and similar levels of performance. 
\section{Conclusion}\label{conclusion}
This paper presented a multi-agent graph convolutional reinforcement learning approach built on top of a weighted fair queuing algorithm with the purpose of meeting stringent demands, in terms of throughput and delay, for a set of classified network flows. The deep learning agents continuously determine the weights with which the flow packets are dequeued. In addition, we implemented two classic multi-agent DQN solutions: one is completely centralized and the other fully distributed. We compare our approaches across different network topologies, scenarios, traffic types, and transport mechanisms, highlighting both their efficiency and the importance of inter-agent communication.\\ \indent These types of solutions are still in their infancy, but as we showed in this work, they can provide promising results. In future works, we will consider a larger scale scenario with multi-layer branches and non-direct connections to the HQ network. We will consider variable neighborhoods dependent on the links between routing nodes. Finally, We will assess our smart queuing proposals alongside a deep reinforcement learning assisted approach to load balancing in networks.
{\small
\bibliography{Library}}
\end{document}